\def\ni{\noindent}
\documentclass{aa}

\usepackage{graphicx}
\usepackage{txfonts}
\usepackage{rotating}
\usepackage{psfig}
\begin{document}

   \title{The VVDS-VLA  Deep Field II.}
   \subtitle {Optical and near infrared identifications of VLA S$_{1.4GHz}>80\mu$Jy
              sources in the VIMOS VLT Deep Survey VVDS-02h field}

   \author{P. Ciliegi \inst{1}, 
G. Zamorani \inst{1}, 
M. Bondi  \inst{2}, 
L. Pozzetti \inst{1}, 
M. Bolzonella  \inst{3}, 
L. Gregorini   \inst{2,4}, 
B. Garilli \inst{5},
A. Iovino \inst{6},
H.J. McCracken \inst{7,8},
Y. Mellier \inst{7,8},
M. Radovich \inst{9},
H.R. de Ruiter  \inst{1},
P. Parma  \inst{2},
D. Bottini \inst{5},
V. Le Brun \inst{10},
O. Le F\`evre \inst{10},  
D. Maccagni \inst{5},
J.P. Picat \inst{12},
R. Scaramella \inst{11},
M. Scodeggio \inst{5},
L. Tresse \inst{10},
G. Vettolani \inst{2},
A. Zanichelli \inst{2},
C. Adami \inst{10},
M. Arnaboldi \inst{9},
S. Arnouts \inst{10},
S. Bardelli  \inst{1},
A. Cappi    \inst{1},
S. Charlot \inst{8},
T. Contini \inst{12},
S. Foucaud \inst{5}
P. Franzetti \inst{5},
L. Guzzo \inst{6},
O. Ilbert \inst{10},
B. Marano     \inst{3},  
C. Marinoni \inst{10},
G. Mathez \inst{12},
A. Mazure \inst{10},
B. Meneux \inst{10},
R. Merighi   \inst{1}, 
P. Merluzzi \inst{9},
S. Paltani \inst{10},
A. Pollo \inst{6},
E. Zucca    \inst{1},
A. Bongiorno \inst{3}, 
G. Busarello \inst{9},
I. Gavignaud \inst{12},
R. Pell\`o \inst{12},
V. Ripepi \inst{9},
D. Rizzo \inst{12}
}

   \offprints{P. Ciliegi; paolo.ciliegi@bo.astro.it}

\institute{ 
INAF-Osservatorio Astronomico di Bologna - Via Ranzani,1, I-40127, Bologna, Italy
\and
IRA-INAF - Via Gobetti,101, I-40129, Bologna, Italy
\and
Universit\`a di Bologna, Dipartimento di Astronomia - Via Ranzani,1,
I-40127, Bologna, Italy
\and
Universit\`a di Bologna, Dipartimento di Fisica - Via Irnerio, 46, I-40126, Bologna, Italy 
\and
IASF-INAF - via Bassini 15, I-20133, Milano, Italy
\and
INAF-Osservatorio Astronomico di Brera - Via Brera 28, Milan,
Italy
\and
Institut d'Astrophysique de Paris, UMR 7095, 98 bis Bvd Arago, 75014
Paris, France
\and
Observatoire de Paris, LERMA, 61 Avenue de l'Observatoire, 75014 Paris, 
France
\and
INAF-Osservatorio Astronomico di Capodimonte - Via Moiariello 16, I-80131, Napoli,
Italy
\and
Laboratoire d'Astropysique de Marseille, UMR 6110 CNRS-Universit\'e de
Provence, Traverse du Siphon-les Trois Lucs, 13012 Marseille, France
\and
INAF-Osservatorio Astronomico di Roma - Roma Italy 
\and
Laboratoire d'Astrophysique de l'Observatoire Midi-Pyr\'en\'ees (UMR 
5572) - 14, avenue E. Belin, F31400 Toulouse, France
}

   \date{Received  October 19, 2004 ; accepted June 16, 2005 }

    \authorrunning{P. Ciliegi et al.}
    \titlerunning{The VVDS-VLA Deep Field. II}

\abstract{
In this paper we present the optical and near-infrared identifications 
of the 1054 radio sources detected in the 20cm deep radio survey 
down to a 5$\sigma$ flux limit of $\sim$80$\mu$Jy  obtained 
with the VLA in the  VIMOS VLT Deep Survey VVDS-02h deep field. 
Using $U,B,V,R,I$ and 
$K$ data,   with  limiting magnitudes of $U_{AB}\sim$25.4,  
$B_{AB}\sim$26.5,
$V_{AB}\sim$26.2,  $R_{AB}\sim$25.9  $I_{AB}\sim$25.0,  $J_{AB}\sim$24.2, 
$K_{AB}\sim$23.9 
 (50\% completeness)   we   identified 718 radio sources
($\sim$74\% of the whole sample).

The photometric redshift analysis shows that, in each 
magnitude bin, the radio sample has a higher median photometric redshift than 
the whole optical sample, while the median $(V-I)_{AB}$ color of the radio sources 
is redder than the median color of the whole optical sample. These results 
suggest that radio detection is preferentially 
selecting galaxies with higher intrinsic optical luminosity. 

From the analysis of the optical properties of the radio sources as function of the radio 
flux,  we found that while about 35\% of the radio sources are 
optically unidentified in the higher radio flux bin (S$>$ 1.0 mJy), 
the percentage 
of unidentified sources decreases to about 25\%  in the faintest bins (S$<$ 0.5 mJy). 
The median $I_{AB}$ magnitude for the total sample of radio sources, 
i.e. including also the unidentified ones, is brighter in the faintest 
radio bins than in the bin with higher radio flux. This suggests that 
most of the faintest radio sources are likely to be associated to relatively
lower radio luminosity objects at relatively modest redshift, rather 
than radio-powerful, AGN type objects at high redshift. Using a classification in 
early-type  and late-type galaxies based on the $(B-I)_{AB}$ color and the photometric redshift,
we found that the majority of the radio sources below $\sim$0.15 mJy are indeed 
late-type star forming galaxies.  Finally,  the radio sources without 
optical counterpart in our deep imaging have a median radio flux of 0.15 mJy, 
equal to that of identified sources.  Given the very faint optical limits, these 
unidentified radio sources probably contain a significant fraction of 
obscured and/or high redshift galaxies. 

    }

   \maketitle

%
%________________________________________________________________

\section{Introduction}

Deep 1.4-GHz counts show an upturn below a few millijansky (mJy), corresponding
to a rapid increase in the number of faint sources. Photometric and 
spectroscopic studies suggest that the faint excess at 1.4 GHz is 
composed predominantly of star-forming galaxies, with a contribution also
from low-power AGNs and early type galaxies. 
However, despite many 
dedicated efforts (see for example Benn et al. 1993, Hammer et al. 1995, 
Gruppioni et al. 1999, Richards et al. 1999, Prandoni et al. 2001) the 
relative fraction of the various populations responsible of the sub-mJy radio 
counts (AGN, starburst, late and early type galaxies), are 
not well established yet. In fact, the photometric and spectroscopic 
work needed in the optical identification process is very 
demanding in terms of telescope time, since a significant fraction 
of faint radio sources have also 
very faint optical counterparts.  It is therefore clear that in order 
to investigate the nature and evolution of the sub-mJy population it is 
absolutely necessary to couple deep radio and optical (both imaging 
and spectroscopic) observations over a reasonably large area of the sky. 

The VVDS-VLA Deep survey is one of the best available surveys to
investigate the nature of the sub-mJy population. A deep radio 
survey has been obtained in  1 deg$^2$ with the VLA  down to a 
5$\sigma$ flux limit of $\sim$80 $\mu$Jy (Bondi et al. 2003). 
A deep multi-color BVRI photometric survey of the whole area is 
available (Le F\`evre et al., 2004a, McCracken et al. 2003), together with
U band (Radovich et al. 2004) and  J and K band (Iovino et al. 2005) data. 
Furthermore, a deep spectroscopic survey is being performed
with the VIMOS spectrograph at the VLT  (Le F\`evre et al. 2004b).

In this paper we present the optical identification 
of the VIMOS radio sources obtained with the photometric data in the 
$U,B,V,R,I$ and $K$ bands.  In Sect. 2 we give a description of the 
available radio, optical and near infrared data, while the description 
of the technique used for the optical identification of the radio 
sources is presented in Sect. 3. Finally, Sect. 4 is a discussion of our 
results, while  Sect. 5 summarizes our conclusion.

\section{The available data}

\subsection{Radio data}

The radio data were obtained with the Very Large Array (VLA)
in B configuration. A 1 square degree mosaic map with an 
approximately  uniform noise
of 17.5$\mu$Jy (1$\sigma$) and with a 6$\times$6 arcsec FWHM gaussian
resolution beam has been obtained. This map
(centered at RA(J2000)=02:26:00 DEC(J2000)=$-$04:30:00)
has been used to
extract a complete catalogue of 1054 radio sources, 19 of which are
considered as multiple, i.e. fitted with at least two separate
component. A detailed description of the radio observations, data reduction,
sources extraction and radio source counts is reported in Bondi et al. (2003).

\subsection{The optical and near infrared data in the $U, B, V, R, I $ and $K$ bands}

Almost the whole square degree of the VVDS-VLA field has been observed
in the $B , V, R $ and $I$ bands with the CFH12K wide-field mosaic camera during
the CFH12K-VIMOS deep imaging survey (Le F\`evre et al. 2004a).
These observations reach limiting magnitudes (50\% completeness for 
point sources) of $B_{AB}\sim$26.5,
$V_{AB}\sim$26.2,  $R_{AB}\sim$25.9 and   $I_{AB}\sim$25.0.
A detailed description  of these optical data is given in
McCracken et al. (2003).
Moreover, a  $U$ band survey has been carried out with the wide 
field imaging (WFI) mosaic 
camera mounted on the ESO MPI 2.2 meter telescope at La Silla, Chile. 
The total effective area covered so far by the $U$ band survey is
$\sim$0.71 deg$^2$. 
The limiting magnitude of the catalogue obtained with these observations 
is 25.4  in $U_{AB}$  (50\% completeness). 
A detailed description of the VIMOS U band 
imaging survey is reported in Radovich et al. 2004.  Finally, a fraction of  the VVDS-VLA 
field has been observed in the $J$ and $K$ bands with the SOFI instrument mounted on 
the ESO NTT  telescope. The total area covered by the $J$ and $K$ bands survey is 
$\sim$ 165 arcmin$^2$, down to  limiting magnitudes of  $J_{AB}\sim$24.2 and $K_{AB}\sim$23.9 
(50\% completeness for point sources).  A detailed description of 
the VIMOS  J and K band imaging survey is reported in Iovino et al. 2005.

In order to obtain a $B_{AB},V_{AB},R_{AB},I_{AB}$ catalogue, all the images 
were combined to obtain a unique detection image
using the $\chi^2$ technique (Szalay et al. 1999).  Subsequently, the photometry in
each bandpass  has been performed at the position defined in the
$\chi^2$ image. The primary advantage in using this technique 
is to simplify the generation of multi-band catalogues and to
reduce the number of spurious detections.  
For a  detailed description  of the  $B_{AB},V_{AB},R_{AB},I_{AB}$  catalogue
see McCracken et al. (2003). Subsequently, the $\chi^2$ image has
been updated with the  $U_{AB}$ , $J_{AB}$ and $K_{AB}$ data (in the areas
covered in these bands) using the same technique described by
McCracken et al. (2003). 

\section{Optical identification}

\subsection{Radio-optical off-set}

\begin{figure}
    \psfig{figure=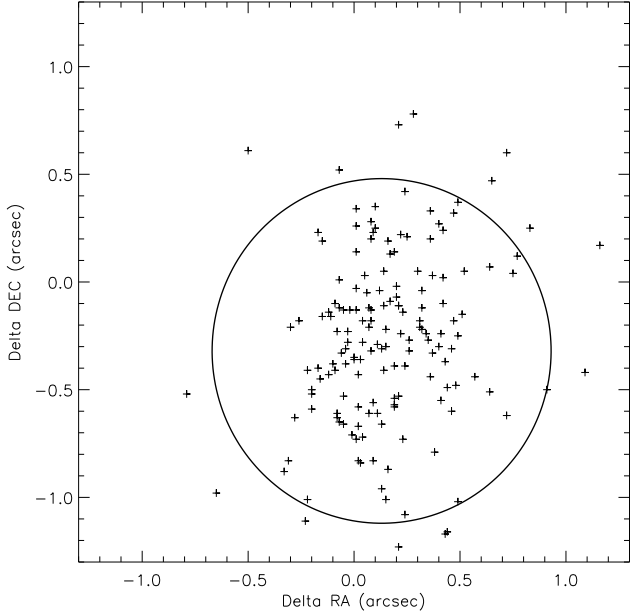,width=9cm,height=8.7cm}
      \caption[]{Position offsets for strong ($>10\sigma$) radio point-like  sources
      identified with a point-like optical counterparts. The circle of radius 0.8 arcsec 
(with the mean offset removed) encloses  90\% of the sources.}
\label{ra_dec_err}
    \end{figure}

The relative off-set between the radio and optical catalogue has been estimated
using the radio position of the 160 unresolved radio sources with a peak flux
density greater than 0.17 mJy ($i.e.$ detected with  peak flux $\sim$ 10$\sigma$)
identified with point like optical counterparts.  Their  $\Delta$RA and
$\Delta$DEC offset are shown in Fig.~\ref{ra_dec_err}. The mean offsets, with their 
standard errors ($\sigma/\sqrt{N}$), are
$<\Delta$RA$>$ = +0.13$\pm$0.03$^{\prime\prime}$ and
$<\Delta$DEC$>$ = $-$0.32$\pm$0.03$^{\prime\prime}$. These offsets should be
removed (subtracting 0.13$^{\prime\prime}$ in RA and adding 0.32$^{\prime\prime}$
in DEC at the radio position) to obtain the radio position in the same
reference frame as the optical CCD.

\subsection{Optical identification of the radio sources}

Of the 1054 radio sources in the complete radio sample, 
57 are outside the presently available catalogues based on
our  CCD data. 
The optical identification of these 57 
sources with the available public catalogues is described in the next Section. 
Moreover, 24 radio sources are within the masked 
regions in the $UBVRIK$ catalogue, since they are spatially coincident 
with bright stars or their diffraction spikes. 
The total number of radio sources for which optical data are  available in the 
 VIMOS CCD  catalogue is therefore 973.

For the optical identification of these 973  radio
sources, we used the likelihood ratio technique, first used in 
this context by Richter (1975) and in modified form by de Ruiter, 
Willis \& Arp (1977), Prestage \& Peacock (1983), 
Sutherland \& Saunders (1992) and  Ciliegi
et al. (2003). The mean off-set between the radio and optical
positions estimated in the previous section has been removed
from the radio positions to compute the positional offset.

The likelihood ratio LR is the ratio between the probability that a given source 
at the observed position and with the measured magnitude is the true optical counterpart, 
and the probability that the same source is a chance background object : 

\begin{equation}
LR=\frac{q(m) f(r)}{n(m)}
\end{equation}

\ni where {\it q(m)}  is the expected distribution as a function of magnitude 
of the optical counterparts, {\it f(r)} is the probability distribution function of the positional
errors, while  {\it n(m)} is the surface density 
of background objects with magnitude {\it m} (see Ciliegi et al. 2003 for a detailed discussion 
on the procedure to calculate {\it q(m)},  {\it f(r)} and {\it n(m)}). 
For each source we adopted an elliptical Gaussian distribution for the
positional errors with the errors in RA and DEC on the radio position reported in the 
radio catalogue (Bondi et al. 2003) and assuming a value of 0.3 arcsec as optical position
uncertainty (McCracken et al. 2003). 

The presence or absence of more than one optical candidates for the same radio source
provides additional information to that contained in {\it LR}.
The reliability $Rel_j$ for object {\it j} being the correct identification
is (Sutherland \& Saunders, 1992)  :

\begin{equation}
Rel_j=\frac{(LR)_j}{\Sigma_i (LR)_i + (1-Q)}
\end{equation}

\ni where the sum is over the set of all candidates 
for this particular source 
and Q is the probability that the optical counterpart of the source is
brighter than the magnitude limit of the optical catalogue 
($Q = \int^{m_{lim}} q(m)~dm $).

The adopted value for $Q$ is 0.65. This value has been estimated by the 
comparison between the expected number of identifications (658) derived from 
the integral of the $q(m)$ distribution  and the 
number of the radio sources (973) that we used in the Likelihood Ratio 
technique (see Ciliegi et al. 2003 for more details). 
However, to check how 
this assumption could affect our results, we repeated the likelihood ratio 
analysis using different values of $Q$ in the range 0.6--0.8. No 
substantial difference in the final number of identifications and in the 
associated reliability  has been found. 

Once {\it q(m)}, {\it f(r)} and {\it n(m)} were obtained,  we 
computed the $LR$ value for all the optical sources within a distance 
of 5 arcsec from the radio position. Having determined the $LR$ 
for all the optical candidates, one has to choose the best threshold 
value for $LR$ ($L_{\rm th}$) to discriminate between spurious and real 
identifications. As  $LR$  threshold we adopted $L_{\rm th}$=0.35. 
With this value, according to Eq. (2) and considering that 
our estimate for Q is 0.65, all the optical counterparts of radio sources 
with only one identification (the majority in our sample)
and $LR>LR_{\rm th}$ have a reliability greater than 0.5. This choice also 
approximately maximizes the sum of sample reliability and completeness. 
With this threshold value we find 718 radio sources with a likely 
identification, 14 of which have two optical candidates with 
 $L_{\rm th}>$0.35 for a total of 732 optical candidates with 
 $L_{\rm th}>$0.35.  The maximum distance between the 
radio and optical position is 2.98 arcsec. The number of expected real 
identifications (obtained by summing  the 
reliability of all the objects with   $L_{\rm th}>$0.35) is about 683,
$i.e$ we expect that about 35 of the 718 proposed radio-optical 
associations may be  spurious positional coincidences. 
In the 14 cases in which more than one optical object with 
$LR>LR_{\rm th}$ has been found associated to the same radio source, 
we assumed the object with the 
highest Likelihood Ratio value as the counterpart of this radio source. 
Among the 255 unidentified radio sources, 17 are empty fields ($i.e.$ 
they have no optical source within 5 arcsec from their position), while 
the other 238  sources have at least one optical source within 5 arcsec, but 
all with $LR < LR_{\rm th}$. 

   \begin{figure}
   \psfig{figure=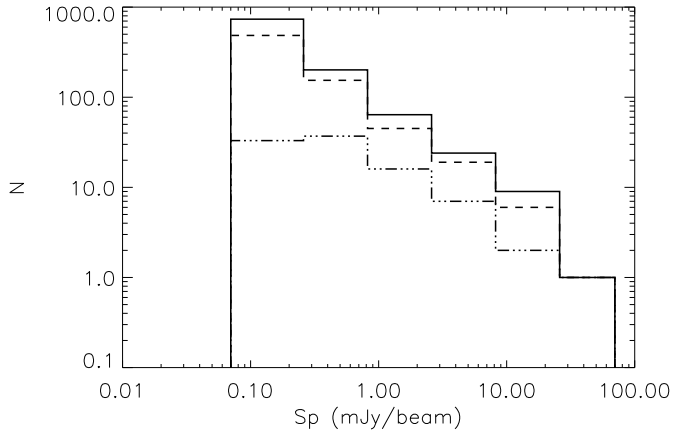,width=9cm}
      \caption[]{Radio peak flux distribution for the whole radio sample (solid line), 
for the 718 radio sources with a reliable optical identification (dashed line) and for 
the 105 radio sources with a near infrared identification in the 2MASS survey 
(dot dot dot dashed line)} 
\label{Speak_id}

    \end{figure}

\subsection{Optical identification of the radio sources outside  the VIMOS CCD catalogue} 

As explained above, 57 radio sources lie outside the currently available VIMOS CCD catalogue. 
For the optical identification of these sources we used the Guide Sky Catalogue 2.2 
\footnote {http://www-gsss.stsci.edu/gsc/gsc2/GSC2home.htm}, 
which provides a two bandpass (J$\sim$B and F$\sim$R) catalogue with a magnitude limit 
of J$\sim$ 19.5 mag  and  F$\sim$ 18.5 mag.  Using a maximum distance of 3 arcsec 
between the radio and optical positions (2.98 arcsec is the maximum 
radio-optical offset for the radio sources identified with the likelihood ratio technique, 
see previous section), we identified 7 radio sources,  only one 
of which has a radio-optical offset greater than 2 arcsec.  The optical properties 
of these 7 radio sources are given in Table~\ref{out_ccd_id}. For each source 
we report the name, the 20 cm total radio flux, the distance between the optical and 
radio position, the J and F magnitude with relative errors and the optical 
classification ${\tt c}$ as reported in the GSC2.2 catalogue (0 for stellar and 3 for 
non stellar).  Given the small number of sources (7 in comparison to the 
718 identified within the CCD area) and the difference in the magnitude systems 
(J and F in the GSC2 catalogue vs. B$_{AB}$ and R$_{AB}$ in the CCD catalogue), 
these 7 sources have not been considered in the analysis and discussion reported
in the next sections and are not included in the identification summary reported in Table~\ref{summary_id}. 

\begin{center} 

\begin{table*}
\centering
  \caption[]{Optical properties of the 7 radio sources outside the CCD area identified with the GSC2.2 catalogue}
  \label{out_ccd_id}
\begin{tabular}{lccccccccc} 
\footnotesize
&&&& \\ \hline
Name & S$_{\rm T}$ & $\Delta$         & J$_{\rm GSC2}$ mag & F$_{\rm GSC2}$ mag & ${\tt c}$ & J$_{\rm 2mass}$  &  H$_{\rm 2mass}$ &   K$_{\rm 2mass}$ \\
     & (mJy)       & ($\prime\prime$) &       &       &       &              &                              \\
\hline 
VIMOS1.4GHz\_J022412-045954  &  0.212  & 2.71 & 14.42  0.34 &   14.59   0.23 &   3 & 16.82 0.19 & 16.14  0.26 & $>$14.78     \\    
VIMOS1.4GHz\_J022526-040116  &  0.709  & 0.21 & 14.15  0.42 &   15.29   0.29 &   3 & 15.26 0.10 & 14.41  0.12 & 13.95  0.09 \\     
VIMOS1.4GHz\_J022549-040023  &  2.125  & 0.41 & 13.47  0.42 &   14.38   0.29 &   0 & 14.74 0.08 & 14.08  0.10 & 13.66  0.08 \\     
VIMOS1.4GHz\_J022603-045902  &  0.682  & 1.49 & 14.01  0.37 &   15.04   0.25 &   3 & 15.36 0.09 & 14.50  0.10 & 14.22  0.08 \\    
VIMOS1.4GHz\_J022604-045932  &  1.079  & 1.57 & 13.01  0.37 &   14.02   0.25 &   3 & 14.86 0.08 & 13.99  0.09 & 13.50  0.07 \\     
VIMOS1.4GHz\_J022609-045804  &  1.792  & 1.28 & 11.79  0.37 &   13.10   0.25 &   3 & 14.14 0.08 & 13.53  0.09 & 12.98  0.07 \\     
VIMOS1.4GHz\_J022628-040051  &  1.257  & 0.16 & 17.03  0.42 &   18.73   0.30 &   3 & 16.19 0.08 & 15.33  0.11 & 14.73  0.11 \\
\hline   

\end{tabular}                                    

\end{table*}

\end{center} 

\normalsize

\begin{table*}
\centering
  \caption[]{Summary of the VVDS-VLA radio sample identification in optical and 
near infrared bands} 
  \label{summary_id}
\begin{tabular}{lcccccccccc} 
&&&& \\ \hline
\multicolumn{10}{c}{\bf Whole radio sample : 1054  sources}  \\ 
&&&& \\ \hline
               &  \multicolumn{7}{c}{\bf Bands}  \\
& & \\
Number of sources :            &  {\bf U}  &  {\bf B} &  {\bf V} &  {\bf R} &  {\bf  I} &  {\bf J} & {\bf K}  & \bf{J$_{\rm 2mass}$} & \bf{H$_{\rm 2mass}$} &  \bf{K$_{\rm 2mass}$}  \\

& & & \\ 

Outside CCD  area                         &      415  &   65      &    66     &   63     &     63      & 989 &   989    & 0    & 0      & 0     \\              
Within masked regions                     &       2   &   24      &    24     &   24     &     24     &   0 &     0    & 0    & 0      & 0     \\ 
With good CCD data                        &     637   &  965      &   964     &  967     &    967     &  65 &    65    & 1054 & 1054   & 1054  \\ 
With reliable identification                       &     371   &  687      &   703     &  714     &    707     &  42 &    43    &  105 &  104   &   93  \\
With reliable identification with I$_{AB}<$24.0   &     361   &  643      &   660     &  668     &    670     &  40 &    40    &  105 &  104   &   93  \\ 

& & & \\

\hline   

\end{tabular}                                    

\end{table*}

\subsection{Near-Infrared identification} 

Only 65 of the 1054 radio sources lie within the 
$\sim$ 165 arcmin$^2$  
covered by our deep K band survey: 21 are unidentified both in the optical 
and K bands, 1 has a reliable optical counterpart but is  unidentified 
in the K band, while  43 have a reliable counterpart both in the optical and 
K bands. To fully explore the near-infrared properties 
of our radio sample, we searched also the Two Micron All Sky Survey 
(2MASS \footnote {http://www.ipac.caltech.edu/2mass/}, Cutri+ 2003 ) 
database for near-IR 
($J ,H$ and $K$ bands) counterparts of our 1054 radio sources.  The 3$\sigma$ limits of 
the 2MASS survey are 17.1, 16.4 and 15.3 mag respectively in the three 
bandpasses $J ,H$ and $K$.  The image pixel scale of the 2MASS detectors is 2.0$^{\prime\prime}$, 
and the positional uncertainties are $\leq$0.5$^{\prime\prime}$.  
 Using a maximum distance of 3$^{\prime\prime}$ between the radio and 2MASS positions, we found 105 reliable 
matches, with 105 detections in $J$, 104 in $H$, and 93 in $K$, including the 7 sources outside 
the optical CCD area reported in Table 1.   
Owing to the relatively shallow flux limits of the 2MASS survey, the surface density of 
background sources is low enough that with a maximum off-set 
of 3$^{\prime\prime}$ between the radio and 2MASS 
positions, chance associations with radio positions are very unlikely. All the 
2MASS sources have $I_{AB}\leq$19.0.  
The  $J ,H$ and $K$ magnitudes have been converted to the AB system using  J$_{AB}$ = J + 0.90, 
H$_{AB}$ = H + 1.37 and K$_{AB}$ = K + 1.84.

\subsection{Identification summary}

A summary of the optical identifications  is reported in Table~\ref{summary_id}. 
For each band we report the number of sources that lie outside the available CCD area, the number of 
sources within the masked regions ($i.e.$ sources coincident with bright stars 
or with their diffraction spikes for which a reliable identification is impossible), 
the number of sources for which reliable data are available (1054 $-$ outside CCD $-$ within masked
regions) and the number of sources with a reliable identification. Finally in the last line 
we report the number of reliable identifications with an $I_{AB}$ magnitude brighter than 24, 
$i.e.$ 
corresponding to the  sources that are potential targets for the VIMOS spectroscopic survey 
(Le F\`evre et al 2004b).  The 7 radio sources identified in the GSC 2.2 catalogue are not included in 
Table~\ref{summary_id}. 
 
In Figure~\ref{Speak_id} we report the radio peak flux distributions for 
the whole radio sample, for the 718 radio sources with a reliable optical 
identification and for the 105 radio sources with a near infrared counterpart 
in the 2MASS survey. As clearly shown in the figure,  while the peak flux 
distribution of radio sources with a 2MASS counterpart is strongly biased towards 
high radio flux (as expected due to the bright limits of the 2MASS survey), the flux 
peak distribution  of the 718 radio sources with an optical identification does not show 
any statistically significant difference with respect to the distribution of the whole radio sample.
The final catalogue of all 718 identified radio sources is available 
on the web at http://virmos.bo.astro.it/radio/catalogue.html. A sample of the catalogue 
is shown in Table~\ref{catalogue}.

\begin{center} 

\begin{table*}
\centering
  \caption[]{Photometric data for the  identified radio sources: a portion is shown here for 
guidance regarding its form and content. The whole catalogue is 
available at the web page  http://virmos.bo.astro.it/radio/catalogue.html} 
  \label{catalogue}
\begin{tabular}{lccccccc} 
\footnotesize
&&&& \\ \hline 
Name & U$_{AB}$ &  B$_{AB}$ &  V$_{AB}$ &  R$_{AB}$ &  I$_{AB}$ &  J$_{AB}$ &  K$_{AB}$ \\

\hline 

 VIRMOS1.4GHz\_J022400$-$040527 &  $>$26.45           & $>$26.35         &  25.52 $\pm$  0.34 &  24.66 $\pm$  0.14 & 23.98 $\pm$ 0.19 & - & - \\
 VIRMOS1.4GHz\_J022400$-$043009 &  $>$26.45           & $>$26.75         &  25.45 $\pm$  0.22 &  24.85 $\pm$  0.19 & 22.76 $\pm$ 0.05 & - & - \\
 VIRMOS1.4GHz\_J022400$-$044950 &   -                 & 20.56 $\pm$ 0.01 &  19.22 $\pm$  0.01 &  18.74 $\pm$  0.01 & 18.15 $\pm$ 0.01 & - & - \\
 VIRMOS1.4GHz\_J022401$-$040734 &  25.61 $\pm$   0.17 & 23.19 $\pm$ 0.04 &  21.71 $\pm$  0.3  &  20.59 $\pm$  0.01 & 19.69 $\pm$ 0.01 & - & - \\
 VIRMOS1.4GHz\_J022401$-$042621 &  24.89 $\pm$   0.12 & 23.94 $\pm$ 0.06 &  23.05 $\pm$  0.05 &  21.94 $\pm$  0.02 & 21.17 $\pm$ 0.02 & - & - \\
 VIRMOS1.4GHz\_J022402$-$040705 &  24.39 $\pm$   0.11 & 24.03 $\pm$ 0.07 &  23.18 $\pm$  0.06 &  23.81 $\pm$  0.09 & 23.04 $\pm$ 0.10 & - & - \\

\hline   

\end{tabular}                                    

\end{table*}

\end{center}

\section{DISCUSSION}

\subsection{Optical magnitude distributions} 

\begin{figure*}
\parbox{9cm}{\psfig{figure=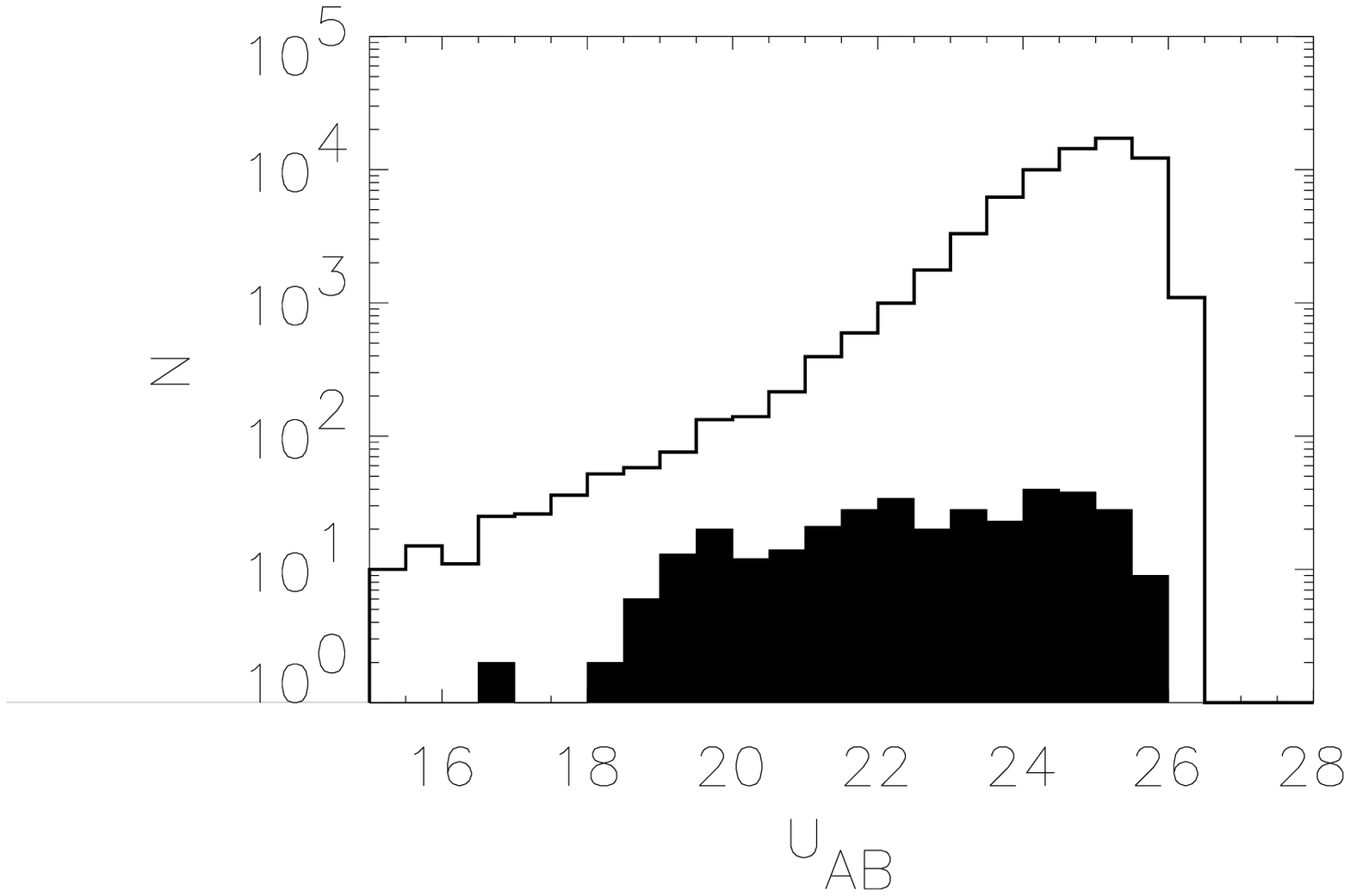,width=9cm,clip=}}
\parbox{9cm}{\psfig{figure=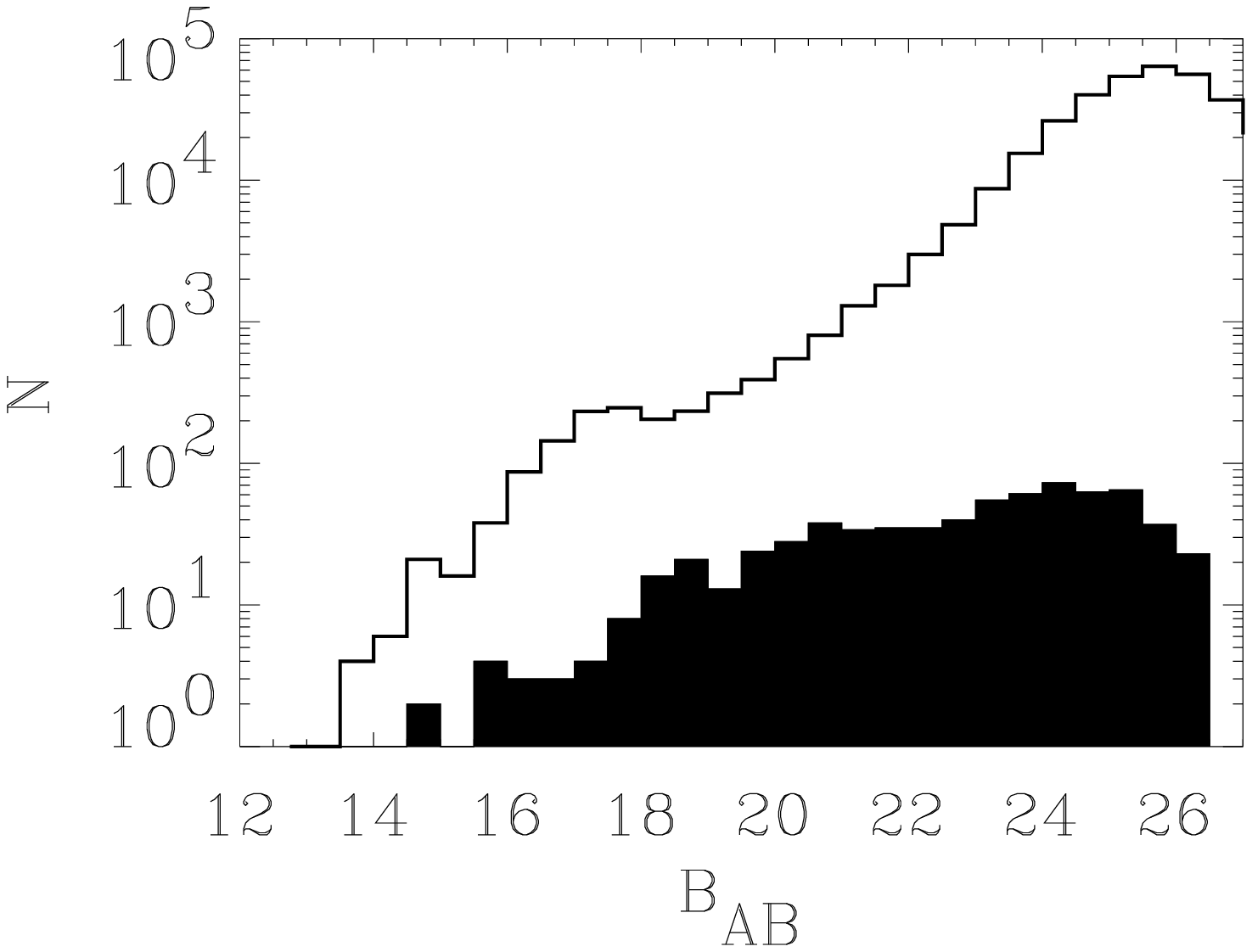,width=9cm}}
\parbox{9cm}{\psfig{figure=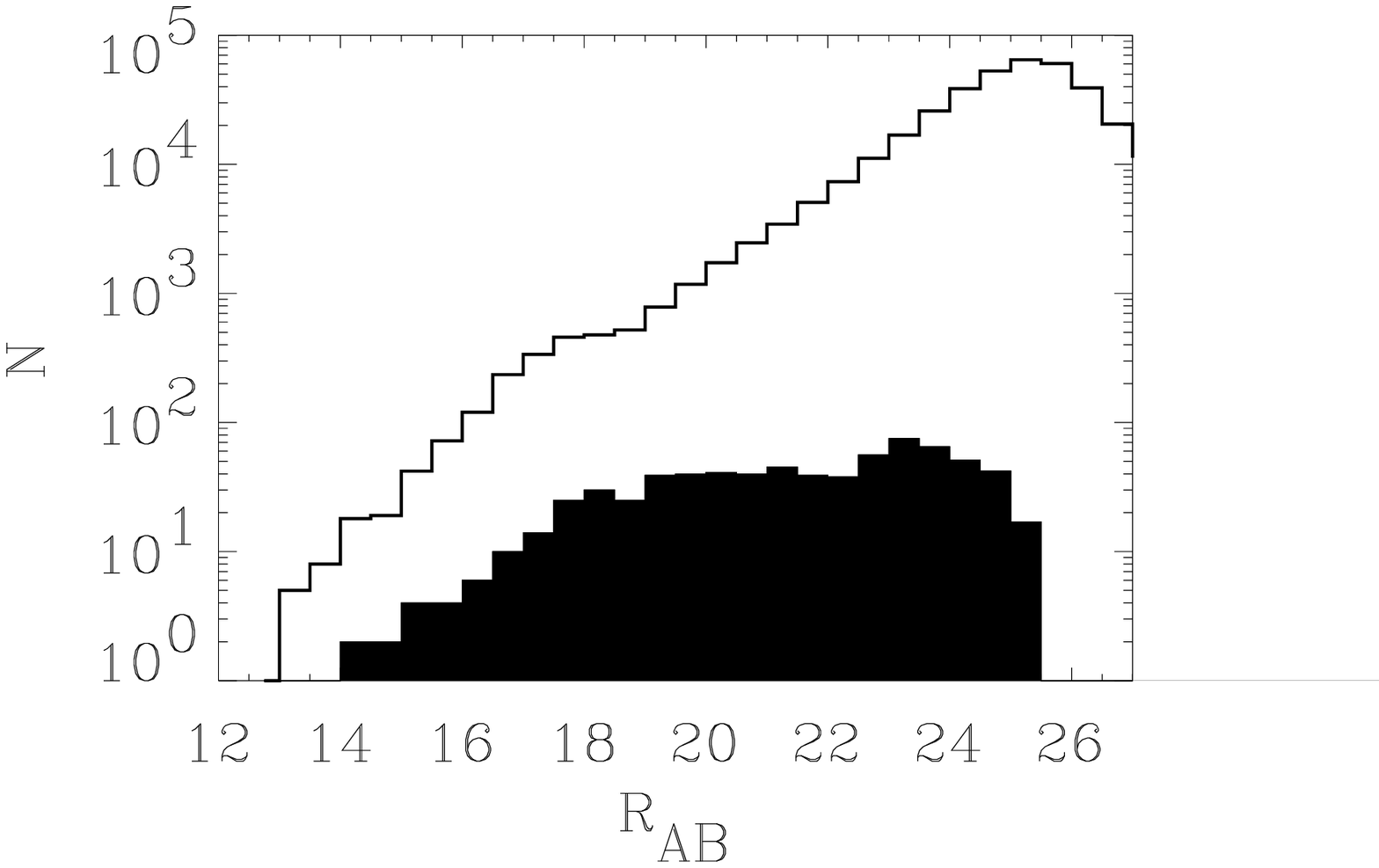,width=9cm,clip=}}
\parbox{9cm}{\psfig{figure=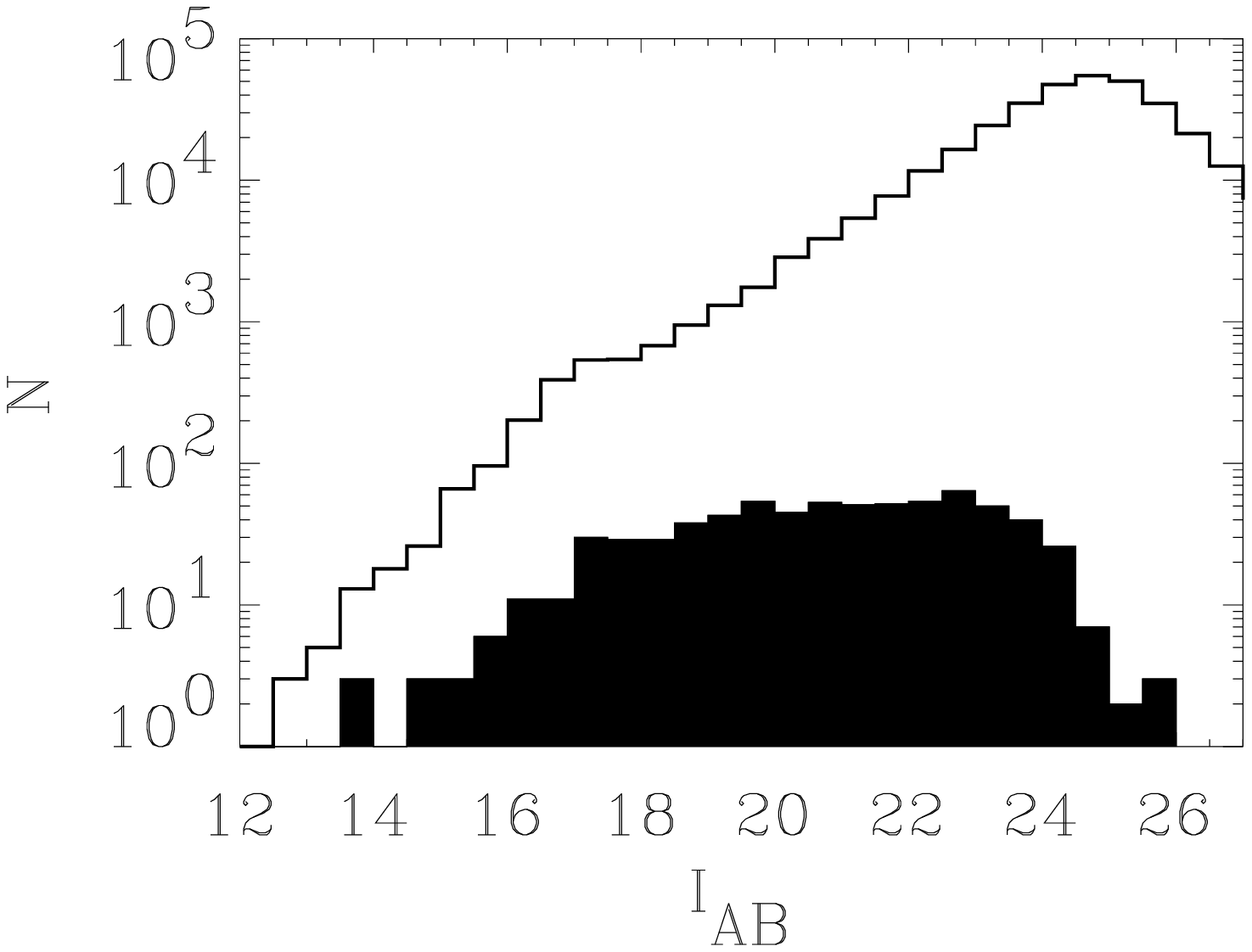,width=9cm}}

\caption[]{Magnitude distributions in the $U_{AB}, B_{AB}, R_{AB}$, and $I_{AB}$ band 
for the 
optical counterparts of the radio sources (filled histograms) and for the whole 
optical sample (empty histograms). } 
\label{mag_dist}

    \end{figure*}

In absence of spectroscopic data, the magnitude and color distributions of the optical 
counterparts can be used to derive some informations on the nature of faint radio sources. 
In Figure~\ref{mag_dist} we show the magnitude distributions in the $U_{AB}, B_{AB}, R_{AB}$, 
and $I_{AB}$ 
bands of the optical counterparts of the radio sources as filled histograms. The empty
histograms show the magnitude distributions of the whole optical data set. 

%%%%%%%%%%%%%%%%%%%%%%%%%%%%%%%
% Magnitude distribution figure
%%%%%%%%%%%%%%%%%%%%%%%%%%%%%%%
%%%%%%%%%%%%%%%%%%%%%%%%%%%%%%%
% End figure
%%%%%%%%%%%%%%%%%%%%%%%%%%%%%%%

This figure clearly shows that the magnitude distributions of the optical 
counterparts of the radio sources are significantly flatter than those 
of the global optical catalogues, reaching a maximum at magnitudes well 
above our optical limiting magnitudes. This turnover in the magnitude 
distribution of faint radio sources was 
initially hinted  in 
the Leiden-Barkeley Deep Survey (LBDS) radio sample (Windhorst 
et al. 1984) and more recently confirmed in the LBDS Hercules subsample 
(Waddington et al. 2000) and  in the identification 
of the radio sources in the Hubble Deep Field region 
(Richards et al. 1999).  

%%%%%%%%%%%%%%%%%%%%%%%%%%%%%%%
% Magnitude I vs 20cm  figure
%%%%%%%%%%%%%%%%%%%%%%%%%%%%%%%

\begin{figure*}
\centering
\psfig{figure=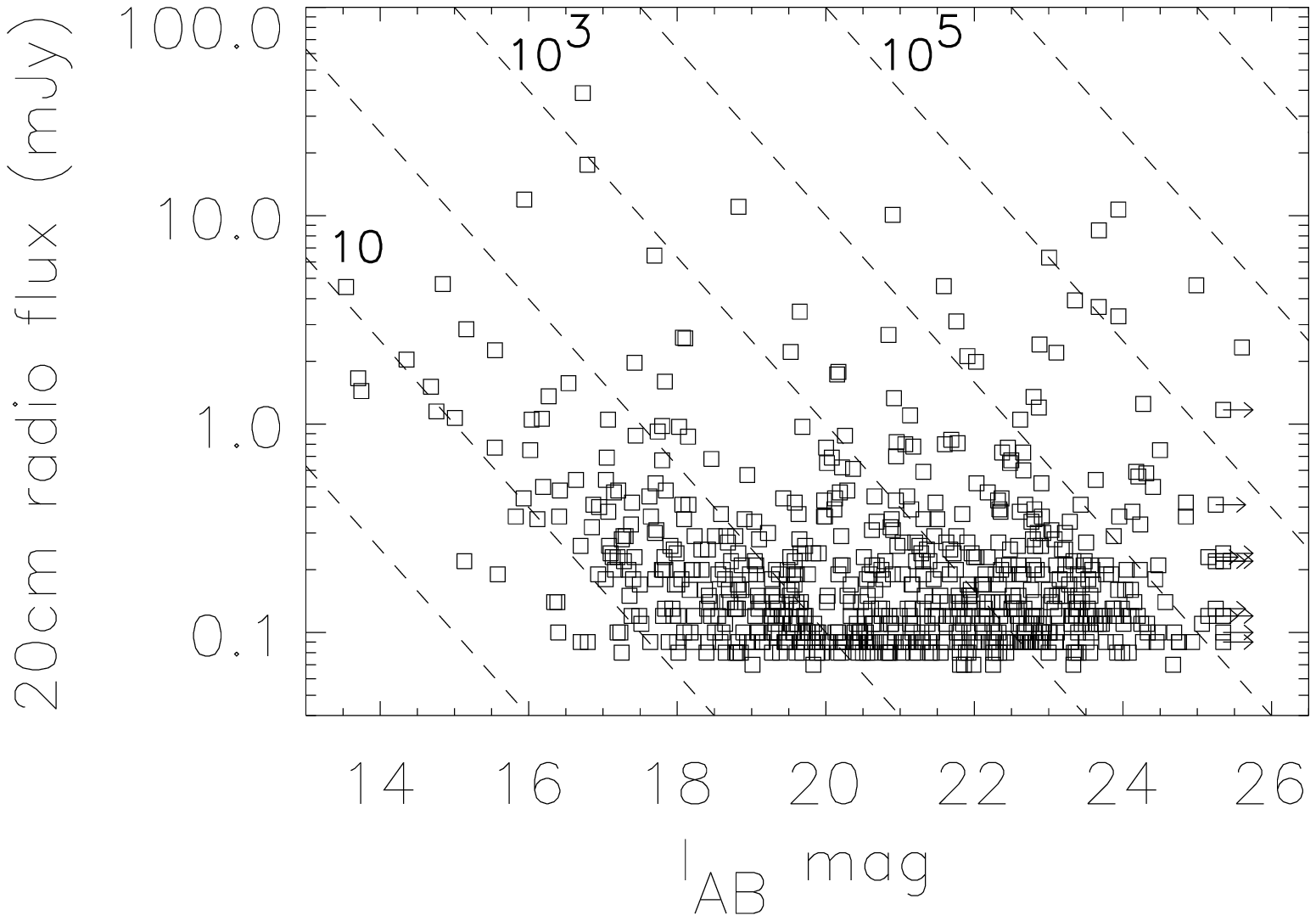,width=15cm}

\caption[]{The $I_{AB}$ band magnitude versus 20 cm radio flux for all the 
radio sources with a reliable optical identification. The lines represent different 
radio to optical ratio $R$, corresponding to $R$=1,10,10$^2$,10$^3$,10$^4$,
10$^5$,10$^6$,10$^7$.  }

%Sources with $R\leq$1000 (expected to be mainly starburst galaxies) are 
%plotted with empty squares while sources with  $R>$ 1000 (expected to be mainly early type 
%galaxies) are plotted with filled squares.  }

\label{Imag_20cm}

\end{figure*}
%%%%%%%%%%%%%%%%%%%%%%%%%%%%%%%
% End figure
%%%%%%%%%%%%%%%%%%%%%%%%%%%%%%%

\begin{figure*}

\centering

\psfig{figure= 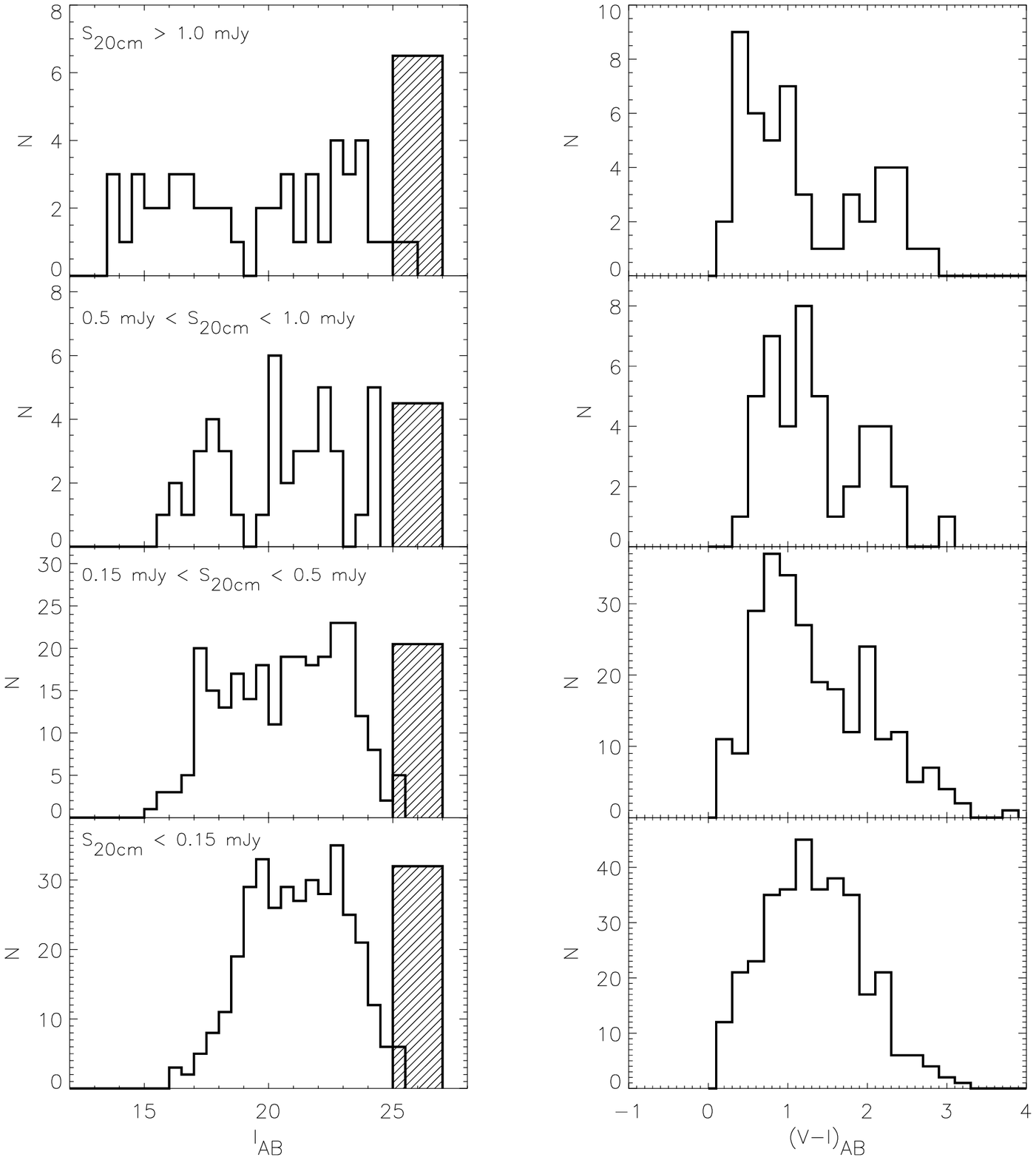,width=16cm}

\vspace{-1.5cm} 

\caption[]{The $I_{AB}$  magnitude (left) and optical color $(V-I)_{AB}$ 
(right) distributions in different radio flux bins for all the radio 
sources with a reliable optical identification. The shaded histograms in the 
left panels represent the unidentified radio sources.} 

\label{histo_radio_bins}

    \end{figure*}

In Figure~\ref{Imag_20cm} we show the 20 cm radio flux versus the $I_{AB}$ 
band magnitude for all the radio sources with an optical identification. 
Superimposed are the lines corresponding to constant values for the observed 
radio-to-optical ratios $\it R$ defined  
as $\it R$ = S $\times$ 10$^{0.4(mag-12.5)}$,
where S and $mag$ are the radio flux in mJy and the apparent magnitude of
the optical counterparts respectively. In 
Figure~\ref{histo_radio_bins} we show the  $I_{AB}$ and $(V-I)_{AB}$ 
distributions in different radio flux bins. 
The shaded histograms in the   $I_{AB}$ magnitude 
distributions show the number of unidentified radio sources, arbitrarily 
plotted in four equal bins between  $I_{AB}$=25.0 and  $I_{AB}$=27.0. 

In Table~\ref{radio_bins}, for each radio flux bin, we report the mean, median 
and standard deviations for the $I_{AB}^{med}$ 
 and  $(V-I)_{AB}^{med}$ color distributions of all the identified radio sources,
the number of radio sources with an optical identification, the total number of 
radio sources (excluding the sources outside the CCD area and the sources in the 
masked regions, see Section 3), the percentage of radio sources with a reliable 
optical identifications and the median  $I_{AB}^{med+unid}$ magnitude calculated considering 
also the unidentified sources, all assumed fainter than  $I_{AB}$=25.0.

%Figure~\ref{Imag_20cm} shows a very mild correlation ({\bf Spearman
%rank correlation coefficient of -0.14 with a deviation 
%of $\sim$3.8$\sigma$ from the 
%non-correlation hypothesis)}, with an extremely 
%large scatter between radio flux and magnitude. 
%This mild correlation, mainly due to objects with small $R$ values, is
%reflected in the change
%with radio flux of the median $I_{AB}$ magnitude ($\Delta I_{AB}^{med} \sim$ 
%1.5 between the brightest and the faintest radio flux bin; see 
%column 2 in Table~\ref{radio_bins}).

Figure~\ref{Imag_20cm} shows  an extremely 
large scatter between radio flux and magnitude, while the analysis of the 
optical properties in different radio flux bins (see Table~\ref{radio_bins}) 
shows a mean (and median) optical magnitude that becomes fainter as fainter 
radio flux bins are considered, although this trend is not statistically 
significant due the large spread of the distributions (see column 4 (standard 
deviations) in Table~\ref{radio_bins}) 

However, while about 35\% of the radio sources are optically 
unidentified in the first  radio flux bin (see Figure~\ref{histo_radio_bins} and
Table~\ref{radio_bins}), the percentage of unidentified sources decreases to about 25\%
in the faintest two radio bins. 
Because of this  decrease of  unidentified sources, although
within the Poisson errors,   
%sources with optical 
%counterpart fainter than $I_{AB} \sim$ 25.0, which suggests a change in 
%the population content around a radio flux S$\sim$ 0.5  mJy, 
%it results that 
the median $I_{AB}$ magnitude for the total sample of
 radio sources, i.e. including also the unidentified ones, is actually
 brighter  of $\sim$0.6 mag  in the
faintest radio bin than in the bin with higher radio flux (see last column in 
 Table~\ref{radio_bins}).  This result shows that the faintest radio
 sources are not in general the faintest sources at optical
 wavelengths and would 
suggest  that most of the faintest radio sources are 
likely to be associated to relatively low redshift star forming objects
with a low radio luminosity, rather than radio-powerful, AGN type objects at high redshift. 
Their median photometric redshift z$_{phot}$ (see next Section for more details on 
 z$_{phot}$) is 0.67, with $\sim$ 90\% of the sources in the photometric redshift 
range 0.1$\leq$z$_{phot}\leq$1.5. 
 
This result is consistent with the expectations from 
previous work in the optical identification 
of the radio sources. Many authors (Windhorst et al. 1995, Richards et al. 1998, 
Richards et al. 1999, Roche et al. 2002) have shown, in fact, that the majority of the 
optical identifications of the $\mu$Jy radio sources are with 
luminous ($L>L_*$) galaxies at relatively modest redshifts (0$\leq$z$\leq$1), 
many of which with evidence for recent star formation. 

Finally, the analysis of the optical color  $(V-I)_{AB}$  in different radio flux bins 
(see  Figure~\ref{histo_radio_bins} and Table~\ref{radio_bins} ) shows that 
there is  a small reddening of the median color as fainter radio fluxes 
are considered.

%%%%%%%%%%%%%%%%%%%%%%%%%%%%%%%
% Magnitude I and V-I color in radio bins 
%%%%%%%%%%%%%%%%%%%%%%%%%%%%%%%

%%%%%%%%%%%%%%%%%%%%%%%%%%%%%%%
% End figure
%%%%%%%%%%%%%%%%%%%%%%%%%%%%%%%

\begin{table*}

  \caption[]{The $I_{AB}$ magnitude (mean, median and standard deviations) and $(V-I)_{AB}$ color 
(mean, median and standard deviations) in four 
different radio flux bins.}
  \label{radio_bins}
\begin{tabular}{lccccccrrrc} 
&& \\ \hline
\footnotesize 
%Radio flux &  $I_{AB}^{med}$ &  $(V-I)_{AB}^{med}$ & N$^{1}_{radio~ID}$ &  N$^{2}_{radio}$ & \% ID & $I_{AB}^{med+unid~~~3}$  \\

Radio flux & \multicolumn{3}{c}{ $I_{AB}$} &  \multicolumn{3}{c}{ $(V-I)_{AB}$} & N$^{1}_{radio~ID}$ &  N$^{2}_{radio}$ & \% ID & $I_{AB}^{med+unid~~~3}$  \\
           & mean & median & $\sigma$    & mean & median & $\sigma$ \\
\hline
\footnotesize
S$_{20cm}\geq$1.00 mJy           & 19.32 & 19.65 & 3.55 & 1.15  & 0.97 & 0.94 &  51 &  77 & 66\% & 22.9 \\  
0.50 $\leq$ S$_{20cm}<$1.00 mJy  & 20.38 & 20.95 & 2.56 & 1.25  & 1.22 & 0.62 &  44 &  62 & 71\% & 22.4 \\
0.15 $\leq$ S$_{20cm}<$0.50 mJy  & 20.59 & 20.89 & 2.31 & 1.29  & 1.18 & 0.76 & 268 & 350 & 77\% & 22.0 \\ 
S$_{20cm}<$0.15 mJy              & 21.14 & 21.18 & 1.90 & 1.25  & 1.26 & 0.71 & 355 & 483 & 74\% & 22.3 \\

\hline   

\end{tabular}

$^{1}$ Number of radio sources with a reliable optical counterpart 

\noindent $^{2}$ Total number of radio sources with good CCD data

\noindent $^{3}$ Median value obtained assuming all the unidentified radio sources with  $I_{AB}>$25.0

\end{table*}

\normalsize

\subsection{Unidentified Radio Sources}

As described in section 3, we have 255  radio sources (26\% of the radio sample) without optical 
counterpart in our optical and near infrared images. The presence of a
significant fraction ($\sim$25-30\%) of radio sources with an optical
magnitude fainter than $\sim$25 was already noted by several authors in
radio surveys of similar depth.  In the identification of the 
 microjansky radio sources in the HDF, only 84 (out of 111,
 $\sim$76\%) sources have been identified 
to $I$=25 mag, with the bulk of the sample identified with relatively 
bright ($I\leq$22) galaxies (Richards et al. 1999), while in the
Phoenix Deep Survey (PDS) about 79\% of the  
sources (659 out of 839) have been identified using  optical $UBVRI$
images to $R\sim$24.5 (Sullivan et al. 2004)

The 255 unidentified VIMOS radio sources have a median 
radio flux of 0.15 mJy, equal to that of identified sources 
(see also Figure~\ref{histo_radio_bins} and Table~\ref{radio_bins} for their 
radio flux distribution). Given the very faint optical counterpart, these 
unidentified radio sources might contain a significant fraction of obscured  
and/or high redshift galaxies. Deep surveys performed at wave-bands free 
from dust absorption ($i.e$ far-infrared and X-ray) may provide the 
most powerful tool to identify these objects.  The VIRMOS VVDS-02h field
is a selected target for far-infrared and X-ray observations with the 
Spitzer and XMM satellites in the framework of the Spitzer Legacy 
Programme SWIRE (Lonsdale et al. 2003) and the XMM Large Scale 
Structure Survey (Pierre et al. 2004) respectively.

\subsection{Photometric Redshifts} 

The optical data in the $B_{AB}$,  $V_{AB}$  $R_{AB}$ and  $I_{AB}$ 
bands, plus $U_{AB}$, $J_{AB}$ and  $K_{AB}$  data (when available) 
 allow us to estimate photometric redshifts for all the 718 
radio sources with an optical counterpart. A detailed description and 
discussion of the method used to estimate the photometric redshift 
in the VIMOS survey is reported in Bolzonella et al. in preparation.

Photometric redshifts have been computed with two codes developed by
the authors: Hyperz, by Bolzonella et
al. (2000) \footnote{http://webast.ast.obs-mip.fr/hyperz/}, using the
Bruzual \& Charlot (2003) library, and the code Le Phare by Arnouts \& Ilbert
\footnote{http://www.lam.oamp.fr/arnouts/LE\_PHARE.html} 
using 72 CWW (Coleman et al. 1980) extended templates.  Both of them have been
modified allowing a training of photometric redshifts using the
spectroscopic data.  

The first step of the method consists in training the photometry: observations
 are affected by possible uncertainties on photometric calibration and zeropoints.  
At the same time, the template SEDs may be not fully representative of the observed 
galaxy population and it may be  difficult to reproduce precisely the response 
functions of filters. Therefore we calibrated photometry and templates with the 
spectroscopic sample in order to obtain a good agreement between the observed 
colour-redshift relation and the one derived from templates. 

Moreover, to avoid spurious
solutions at high redshift, frequently found when  the U magnitude
is not available, we imposed a prior in the redshift distribution.  To
this aim we used the formalism described in Ben\'{\i}tez (2000),
without considering the mix of spectrophotometric types.  We applied a
prior roughly reproducing the $N(z)$ of the spectroscopic sample, but
with a non null probability of being at high redshift.  
%In this way
%we suppress high probability peaks at high redshifts, continuing to
%keep them as possible solutions.  
By imposing this prior we do not
significantly affect the redshift distribution and we considerably improve
the agreement between photometric and spectroscopic redshifts, in
particular when only the optical magnitudes are available.  The values
of photometric redshifts used in this paper have been obtained with the
second of the two codes, Le Phare, although the two codes produced very
similar results.

%With regard to the radio sample, only 45 objects over 718 have
%photometric data spanning from U to K, therefore the application of
%the technique explained above allow us to obtain reliable redshift
%estimate for the whole sample.

So far, only a small fraction (54/718) of the optical counterparts of
the radio sources has a spectroscopically measured redshift.  For these
sources the spectroscopic and photometric redshifts are in good
agreement (90\% of the objects have $\Delta z/(1+z_{\rm spec}) <
0.07$). The analysis of the sub-sample of spectroscopically identified
radio-sources is in progress and will be presented elsewere.

\begin{figure*}

\centering

\psfig{figure=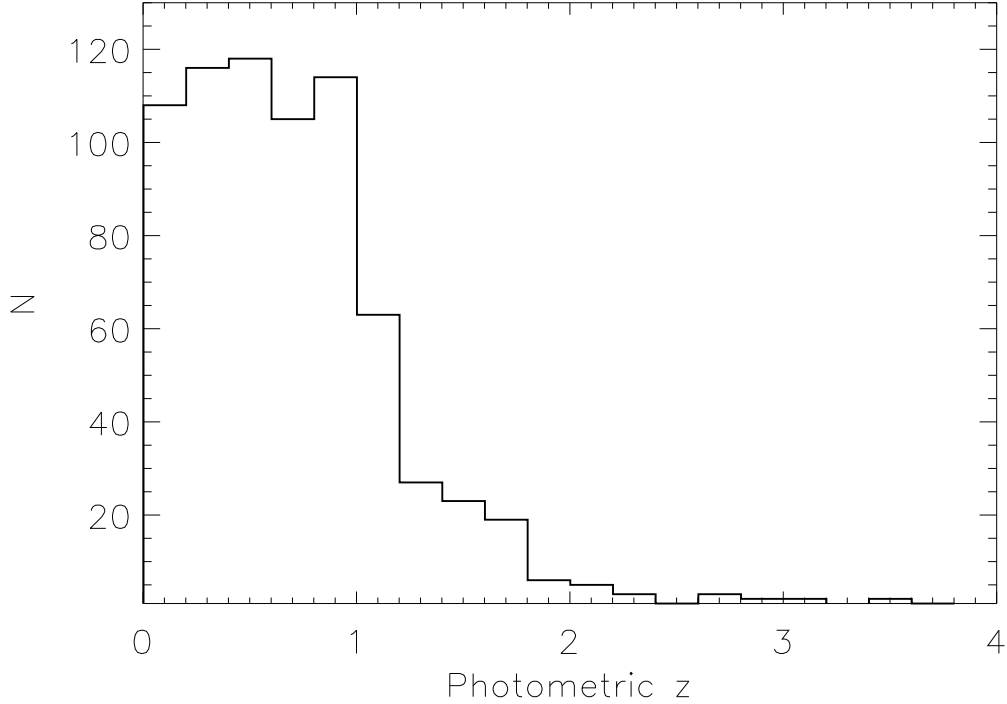,width=15cm}

\caption[]{ The photometric redshift histogram for the 718 radio 
sources with a reliable optical counterpart} 

\label{z_phot}

    \end{figure*}

Figure~\ref{z_phot} shows the photometric redshift histogram for all 
the radio sources with a reliable optical counterpart.  About 
80\% of the sources are estimated to be at z $\leq$ 1, 
with a small high redshift tail extending up to z $\sim$ 3,  although 
among the 157 radio sources with z$_{phot}>$1.0 we have the $J$ and $K$ 
bands data (and then a more reliable   z$_{phot}$) for only 11 sources. This redshift 
distribution is in good agreement with that obtained by Sullivan et al. 2004 in their 
analysis of the optical and near infrared counterparts of the radio sources in the 
Phoenix Deep Survey.

%%%%%%%%%%%%%%%%%%%%%%%%%%%%%%%
% Photometric redshift histogram 
%%%%%%%%%%%%%%%%%%%%%%%%%%%%%%%

%%%%%%%%%%%%%%%%%%%%%%%%%%%%%%%
% End figure
%%%%%%%%%%%%%%%%%%%%%%%%%%%%%%%

\subsection{Color-color diagrams and comparison with the whole optical data set} 

%%%%%%%%%%%%%%%%%%%%%%%%%%%%%%%
% color-color plots
%%%%%%%%%%%%%%%%%%%%%%%%%%%%%%%
\begin{figure*}
\parbox{9cm}{\psfig{figure=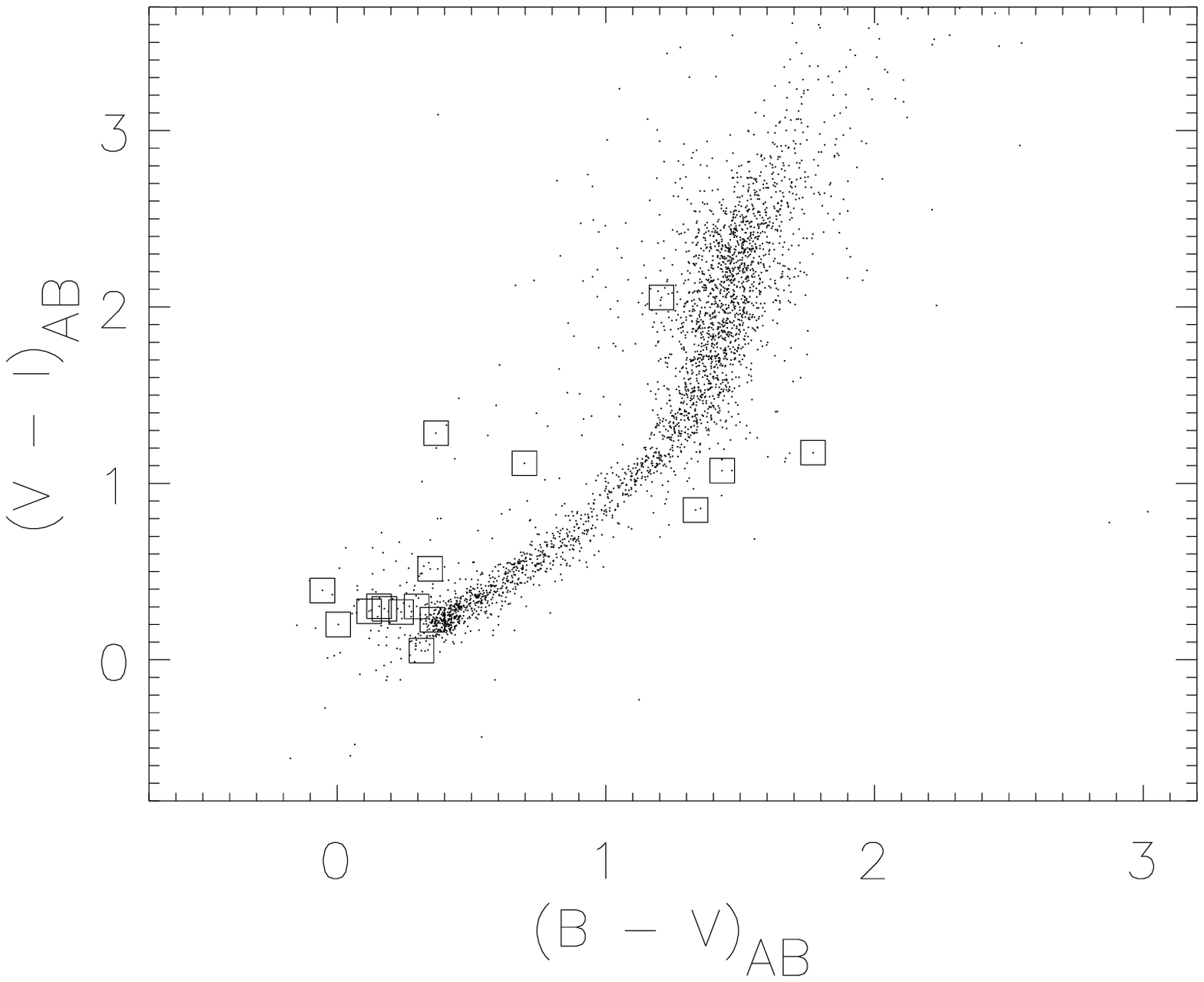,width=9cm}}
\parbox{9cm}{\psfig{figure=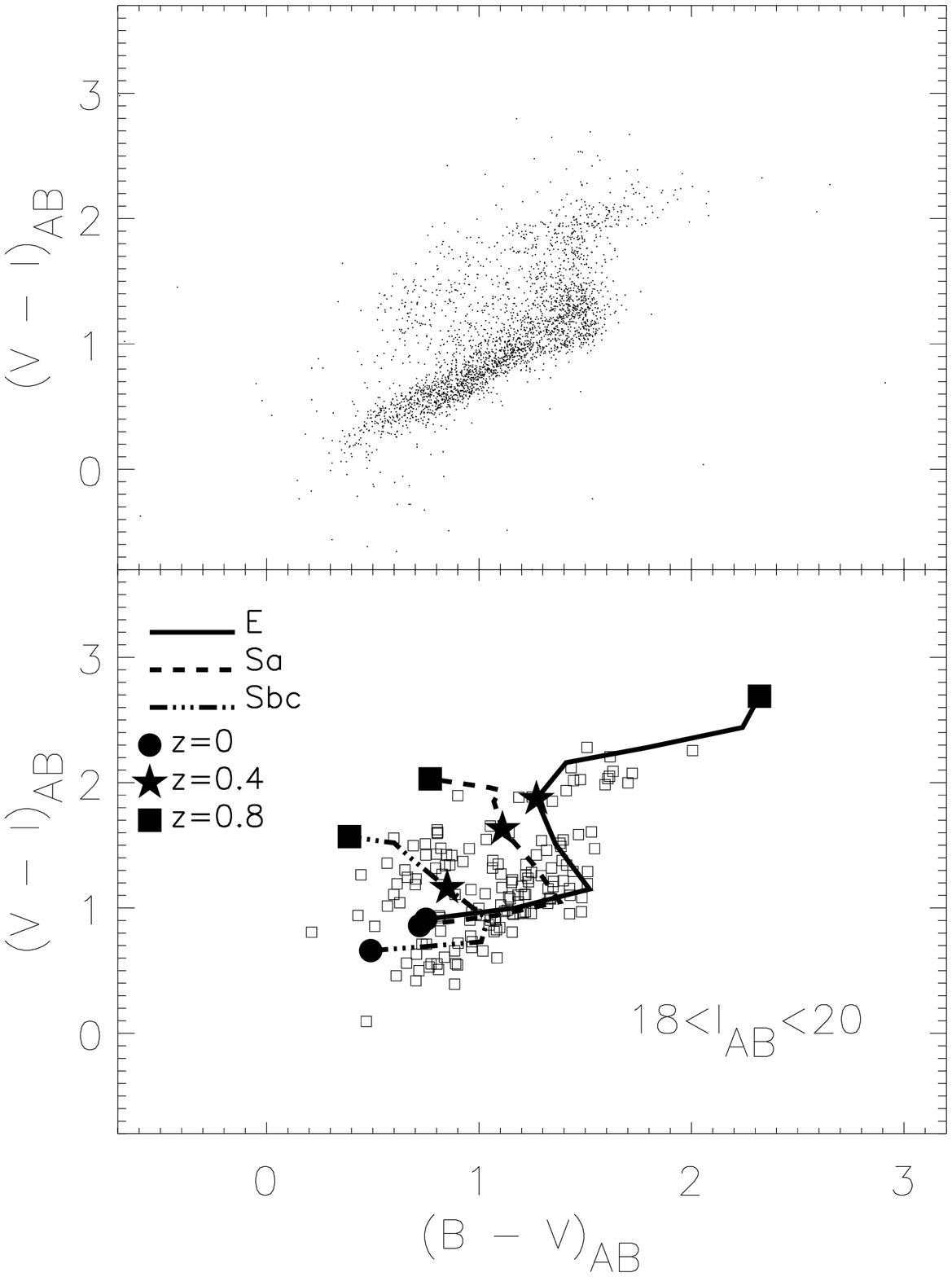,width=9cm}}
\parbox{9cm}{\psfig{figure=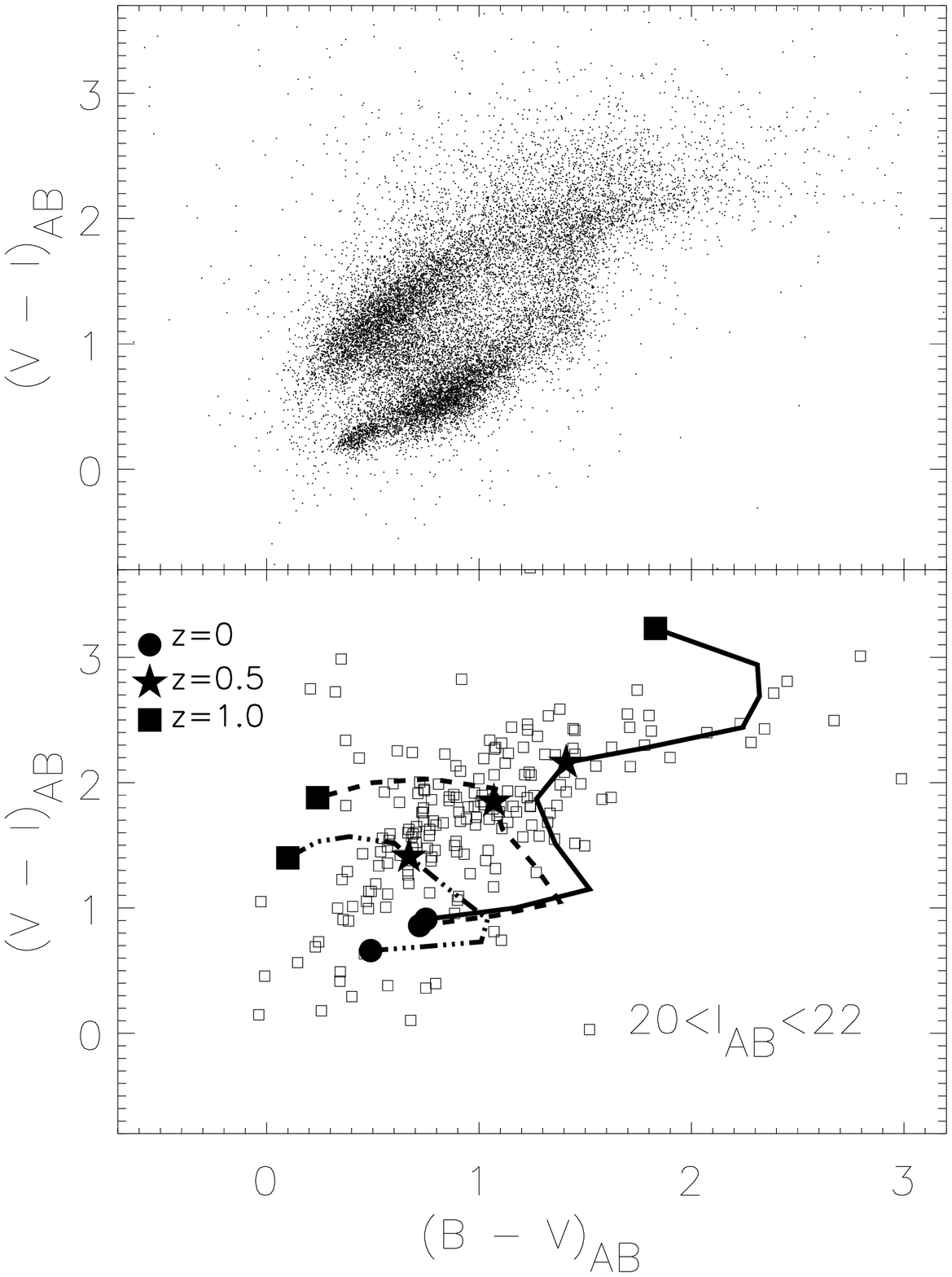,width=9cm,clip=}}
\parbox{9cm}{\psfig{figure=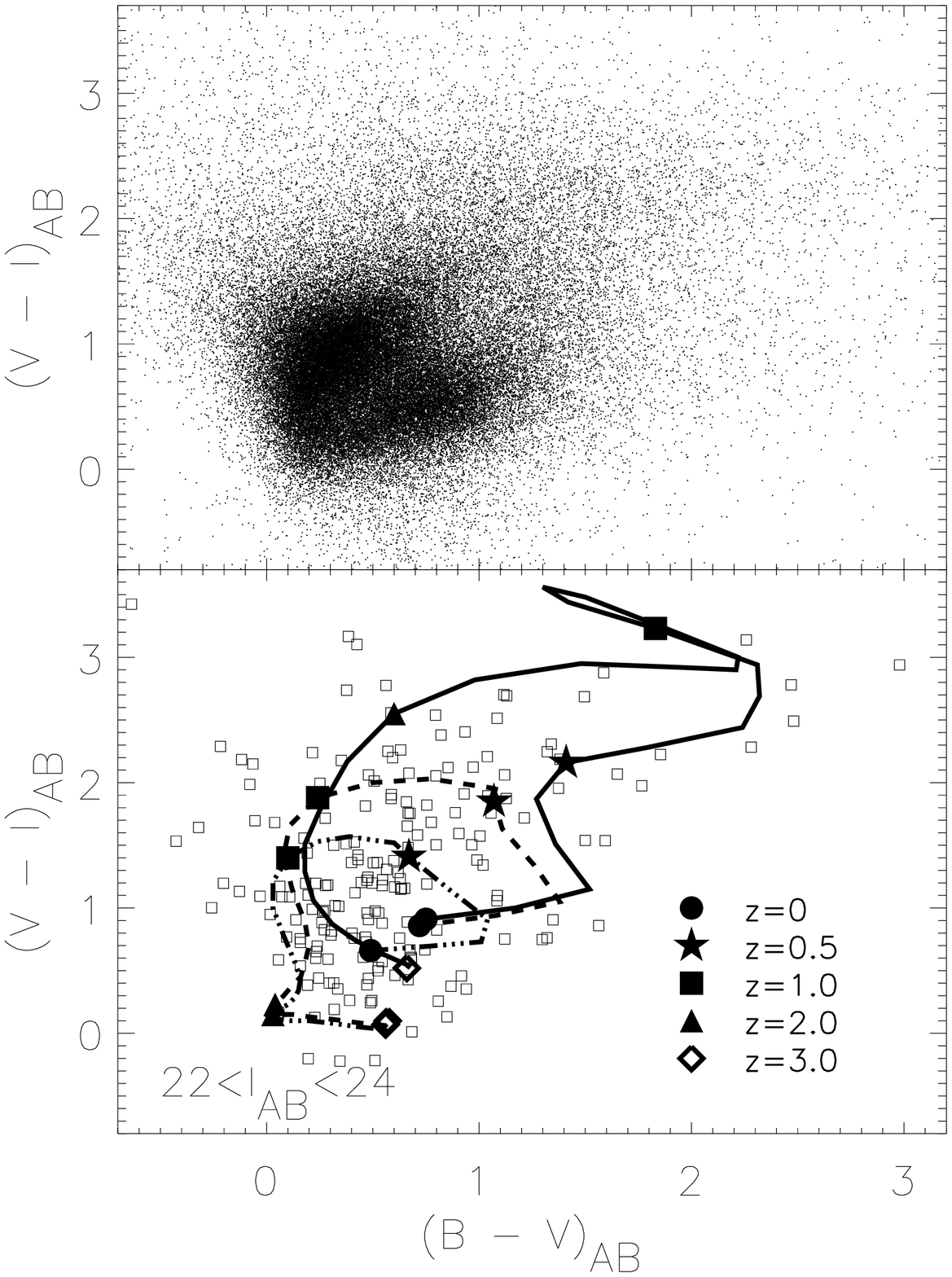,width=9cm}}

\caption[]{$(B-V)_{AB}$ and $(V-I)_{AB}$ colors for the whole data 
set (small dots) and for the optical counterparts of the radio sources
(open squares).  
The sources classified as 
point-like in the optical band ($I_{AB}<$21.5) are plotted in the top-left panel, while the 
sources classified as extended are plotted in the other three panels 
in three different magnitude ranges. Solid,  dashed and dashed dot-dot-dot lines 
show the path in redshift of early-type, Sa and Sbc galaxies. 
The evolutionary tracks were computed  using the ``2000'' 
revision of the GISSEL libraries (Bruzual \& Charlot, 1993). The model tracks span the 
redshift range 0.0$<z<$0.8 for the magnitude slice 18$<I_{AB}<$20, 
0.0$<z<$1.0  for the magnitude slice 20$<I_{AB}<$22 and 
0.0$<z<$3.0  for the magnitude slice 22$<I_{AB}<$24. } 

\label{color_color}

    \end{figure*}
%%%%%%%%%%%%%%%%%%%%%%%%%%%%%%%
% End figure
%%%%%%%%%%%%%%%%%%%%%%%%%%%%%%%

In this section we compare the optical colors properties of the radio sources 
with those of the whole optical data set.  
A detailed discussion of the color properties of the whole 
optical data set with a comparison to those from other deep surveys 
is reported in McCracken et al. (2003). 

In Figure~\ref{color_color} 
we show the $(B-V)_{AB}$ and $(V-I)_{AB}$ colors for the whole data 
set (small dots) and for the optical counterparts of the radio sources 
(open squares). The sources classified as point-like in the optical band 
($I_{AB}<$21.5) are plotted in the top-left panel, while the 
sources classified as extended are plotted in the other three panels 
in three different magnitude ranges (18$<I_{AB}<$20, 20$<I_{AB}<$22 and 
22$<I_{AB}<$24, respectively). Solid, dashed and dashed dot-dot-dot lines 
show the path in redshift of early-type, Sab and Sbc 
galaxies. The evolutionary tracks were computed using the ``2000'' 
revision of the GISSEL libraries (Bruzual \& Charlot, 1993). 
The median value of   $(B-V)_{AB}$, $(V-I)_{AB}$, z$_{phot}$ and $I_{AB}$
for the 
sources classified as extended are reported in Table~\ref{color}, where the same values
are given also for the whole optical sample in the same magnitude
ranges. The errors on the median 
values reported in Table~\ref{color} have been calculated using
1.2533$\sigma$/$\sqrt{N}$,  where $\sigma$ is the 
standard deviation of the distribution and N is the number of objects (Akin \& Colton 1970). 

Moreover, for each data set ($(B-V)_{AB}$, $(V-I)_{AB}$, z$_{phot}$ and $I_{AB}$) in 
each magnitude bin, the Kolmogorov-Smirnov test (KS) has been used to test the hypothesis 
that the two distributions (that from the whole optical sample and that from the radio 
sources) are drawn from the same distribution. The results of the KS test are reported 
in  Table~\ref{color} (last row in each magnitude bin). In particular, for each KS test performed, 
we report the significance level of the KS statistic. Small values show that the two distributions 
are significantly different. 

\begin{table*}

\centering

  \caption[]{The median values for $(B-V)_{AB}$, $(V-I)_{AB}$,  
photometric redshift z$_{phot}$ and $I_{AB}$ magnitude in three different magnitude slices 
for the whole optical data set and for the optical counterparts of the radio sources.  
The last column  reports 
the number of sources in the magnitude bin. }
  \label{color}
\begin{tabular}{lccccr} 
&& \\ \hline
Sample &  $(B-V)_{AB}$ &  $(V-I)_{AB}$ & z$_{phot}$  & $I_{AB}^{med}$ & N \\

\hline

&& \\
\multicolumn{6}{c}{\bf Magnitude range  18$<I_{AB}<$20} \\

&& \\

Optical sample                 & 1.10 $\pm$ 0.01 & 1.06 $\pm$0.03 & 0.31 $\pm$ 0.01  & 19.31$\pm$0.01 & 3056 \\  
Radio sources                  & 1.08 $\pm$ 0.03 & 1.17 $\pm$0.04 & 0.39 $\pm$ 0.02  & 19.22$\pm$0.05 &  153 \\
KS test significance           & 0.13            & 0.005          & 0.04             & 0.04          &      \\

&& \\
\multicolumn{6}{c}{\bf Magnitude range  20$<I_{AB}<$22} \\

&& \\

Optical data set               & 0.82 $\pm$ 0.01 & 1.16 $\pm$0.01 & 0.56 $\pm$ 0.01 & 21.32$\pm$0.01  & 17962 \\  
All radio sources              & 0.93 $\pm$ 0.03 & 1.77 $\pm$0.06 & 0.82 $\pm$ 0.04 & 21.03$\pm$0.05 &   190  \\
KS test significance           & 0.003           &$<1\times10^{-8}$& $<1\times10^{-8}$ & $3.6\times10^{-6}$ & \\

&& \\
\multicolumn{6}{c}{\bf Magnitude range  22$<I_{AB}<$24} \\

&& \\

Optical sample                 & 0.46 $\pm$ 0.01 & 0.88 $\pm$0.01 & 0.80 $\pm$ 0.01 & 23.33$\pm$0.01  & 82944\\  
Radio sources                  & 0.53 $\pm$ 0.05 & 1.22 $\pm$0.07 & 0.96 $\pm$ 0.05 & 22.87$\pm$0.05  & 206 \\
KS test significance           & 0.03            &$<1\times10^{-8}$& $<1\times10^{-8}$ & $<1\times10^{-8}$ & \\

\hline   

\end{tabular}                                    

\end{table*}

The color-color plot for point-like sources (top-left panel of  Figure~\ref{color_color}) 
shows that, as expected, the majority of the radio sources with a point-like optical 
counterpart have different colors with respect to the bulk of the whole optical data 
set. In fact, while the majority of the  point-like sources are 
expected to be stars, those associated with radio sources are, with very high 
probability, active galactic nuclei (AGN) or actively star forming compact galaxies. 

The top-right panel of   Figure~\ref{color_color} shows the bright 
magnitude slice  at 18$<I_{AB}<$20 for extended sources.  
As shown in the Figure and in Table~\ref{color} the color properties 
of the radio sources in this bright magnitude 
slice are not significantly different from those of the whole optical data set.  
The bulk of the 
optical and radio sources are consistent with model tracks of a galaxy population 
around z$\sim$ 0.2-0.3 but the median photometric redshift  of the 
radio sources is  higher than that of the whole optical sample  (see
Table~\ref{color}). 

The lower-left panel of  Figure~\ref{color_color} shows the intermediate
 magnitude slice at 20$<I_{AB}<$22.  As already noted by McCracken et al. (2003), 
the whole galaxy population (predominantly late-type galaxies) occupies two distinct loci, 
and in this color-color space there is a reasonably well defined separation between high 
(z$>$0.4) and low (z$<$0.4) redshift galaxies. However, as clearly shown in the figure, 
the radio sources essentially populate only the locus  of the high redshift (z$\sim$0.4-0.5) galaxies.
This  is  confirmed by the photometric redshift 
analysis, with a median redshift  of the whole data sample (0.56) significantly lower than the median 
redshift of the radio sample (0.82; see Table~\ref{color}).  
The median $(V-I)_{AB}$ color 
of the radio sources ($(V-I)_{AB}^{med}$=1.77, see  Table~\ref{color}) is also significantly redder 
than that of the whole optical data set ($(V-I)_{AB}^{med}$=1.16).  The KS test shows 
that the $(V-I)_{AB}$, z$_{phot}$ and $I_{AB}$ distributions of the whole optical sample 
are very significantly different from that of the radio sources (see Table~\ref{color}). 

Finally, in the fainter magnitude slice (22$<I_{AB}<$24, lower-right panel of  
Figure~\ref{color_color}) the colors are bluer than in the brighter magnitude slice
both for the radio sources and the whole optical data set. Even if less evident from 
the figure, also in this magnitude slice there is a statistically significant 
difference between the color, photometric redshift and magnitude  distributions 
of the radio sources and the radio quiet galaxies  (see results of KS test in 
 Table~\ref{color}). 

The fact that, in each magnitude bin, the median photometric redshift of the 
radio sample is higher than that of the whole optical sample shows that 
radio detection is preferentially selecting galaxies with higher intrinsic 
optical luminosity. This conclusion is strengthened by the fact that, because 
of the different magnitude distributions (see Figure 3), the median magnitude $I_{AB}^{med}$ 
of the radio sample is brighter than that of the optical sample in each magnitude 
bin (see  column 5 in Table~\ref{color}).  

\subsection{Color - redshift diagram for the radio sample} 

\begin{figure*}
\centering
\psfig{figure=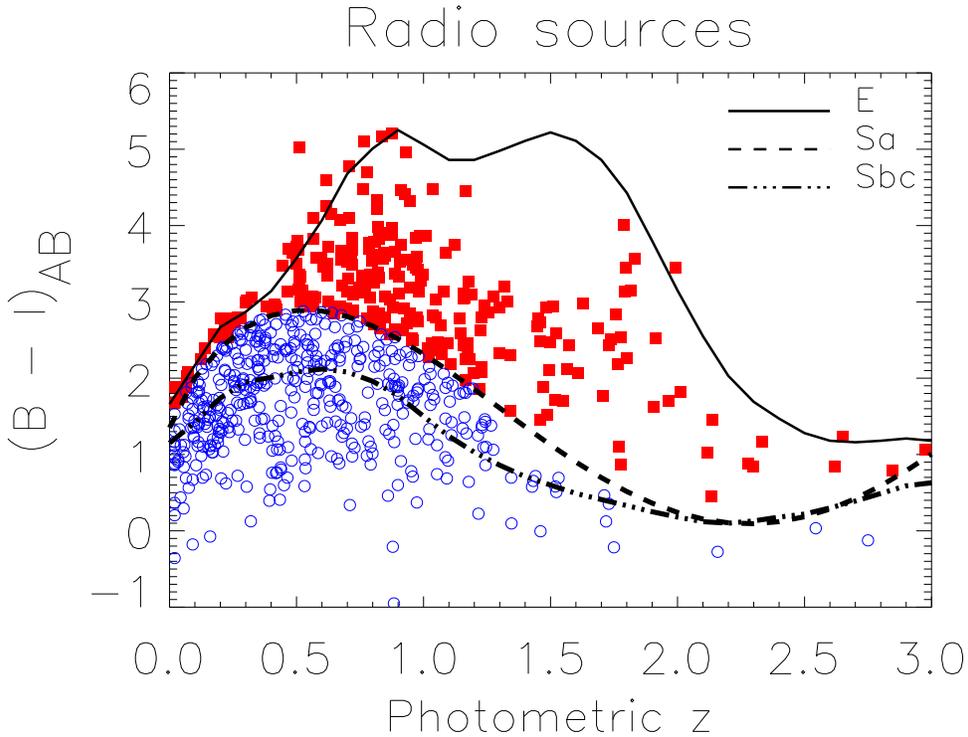,width=15cm}

\caption[]{The $(B-I)_{AB}$ color versus the photometric redshift for the  radio 
sources with a reliable optical counterpart. Sources above the Sa track (expected to be 
mainly early type galaxies at z$\geq$0.5) are plotted with 
filled squares, while the sources below the Sa track (expected to be mainly late-type, 
star-forming galaxies at all redshift) are plotted  with empty circles (see text for more 
details). } 

\label{z_phot_BI_sa}

    \end{figure*}

In Figure~\ref{z_phot_BI_sa} we show the optical color $(B-I)_{AB}$ versus 
the photometric redshift for all the radio sources with an optical counterpart, 
together with predicted loci for early type galaxies (solid line), Sa galaxies (dashed line) 
and Sbc galaxies (dashed dot-dot-dot line).  Sources above the Sa track are expected to be 
mainly early type galaxies and most of them  (222/263)  
have z$_{phot}\geq$0.5, while the sources below 
the Sa track are expected to be mainly late-type, star-forming galaxies and they more 
uniformly occupy the entire redshift range, including the low redshift one. 

In order to study, at least statistically, the properties of the two classes of objects, 
we have adopted in the following the Sa track as separation between early and late type
galaxies. We are fully aware that such a classification is only a first approximation, 
which does not work on an object-by-object basis, and can be significantly uncertain 
especially at redshift greater than $\sim$1. In fact, above z$\sim$ 1 the photometric 
redshifts are more uncertain. This, coupled with the rapidly decreasing 
expected $(B-I)_{AB}$ color of the Sa track up to z$\sim$2, makes the separation between 
early and late type galaxies only indicative in this redshift range. 
A somewhat similar separation has been already used by Ciliegi et al. 
2003 in their analysis of the 6cm radio sources in the Lockman Hole. They used the color 
V-K=5.2 to separate the high redshift (z$\geq$0.5) early-type galaxies from the late-type,  
star forming galaxies at all redshifts. Due to the lack of the K-band covering for the 
whole VIMOS radio sample, we cannot use the V-K color to separate the two classes. 
However, as clearly shown in their Figure 13, the two methods are  largely equivalents: 
only two sources (out of 63 objects) have a V-K color consistent with early-type 
galaxies (V-K$>$5.2)  but lie well below the Sa track. 

\begin{figure*}
\centering 
\psfig{figure=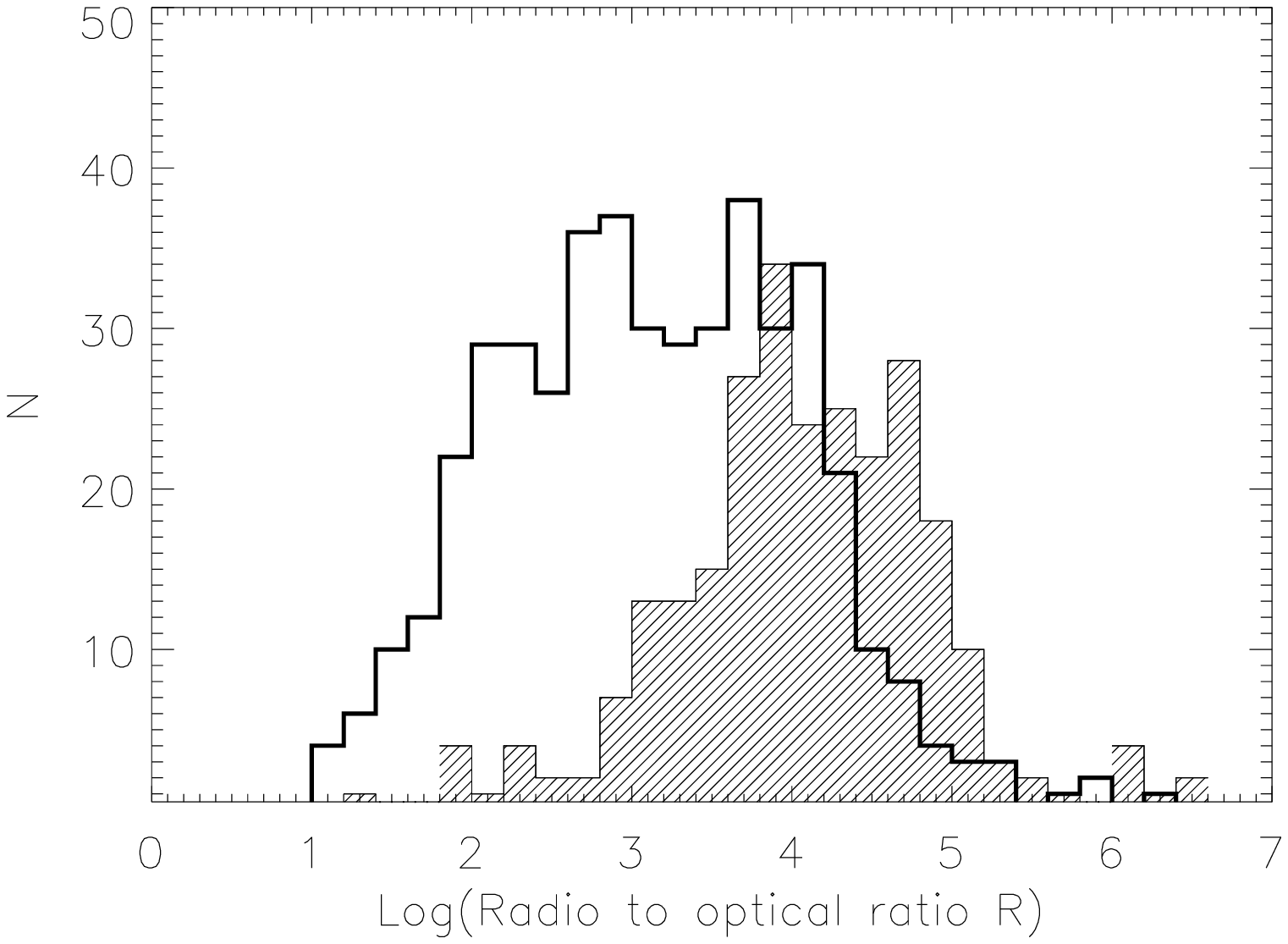,width=15cm}

\caption[]{Distribution of the radio-to-optical ratio $R$ for the sample of early type galaxies 
(shaded histogram) and for the sample of late type galaxies (empty histogram). } 

\label{Ratio_early_late}

    \end{figure*}

We then used these two  radio sub-samples selected on the basis of the Sa track 
to search for possible differences in the radio-to-optical ratio R (defined as in Section 4.1) 
between  early and late type galaxies.  
Previous works on the optical identifications of radio sources 
(Kron et al. 1985, Gruppioni et al. 1999, Richards et al. 1999, Geogakakis et al. 1999, 
Prandoni et al. 2001, Ciliegi et al. 2003) have suggested, in fact, that  
the majority of  the sources with low radio-to-optical 
ratio  are associated with star forming galaxies, characterized by moderately weak intrinsic 
radio power. Viceversa, early type galaxies, in which radio luminosity is likely connected 
to nuclear activity,  cover a much larger range in radio power, and hence in $R$, 
becoming the dominant population at high $R$.  In Figure~\ref{Ratio_early_late} we present 
the distribution of the radio-to-optical ratio $R$ for  the two samples of early type galaxies 
(shaded histogram) and the sample of late type galaxies (empty histogram).  
The two distribution are very significantly different (at more than 7$\sigma$ level, on the basis of KS test):   late-type, star forming 
galaxies are the dominant population at low radio-to-optical ratio $R$ (median 
Log$R$ = 3.08 $\pm$ 0.05, using the $B_{AB}$ magnitude to calculate $R$), while 
the majority of the galaxies in the region of 
high $R$ are early-type galaxies (median Log$R$= 4.06 $\pm$0.06). 
In the high R tail we still have a component of objects classified as late type galaxies 
(about one third of the objects with Log$R\geq$ 4.6). 
A better understanding of the physical nature of these 
objects, which may include also star forming galaxies in which the high 
$R$ value is due to significant dust obscuration, requires spectroscopic data.

Finally, we estimated  the relative fraction of the two 
populations (early and late type galaxies) in different radio flux bins. 
We used the four flux bins defined  in Table~\ref{radio_bins}.
Assuming, consistently with Figure~\ref{Ratio_early_late}, that about 2/3 of 
the unidentified radio sources (which have, by definition,
high $R$ values) are early-type galaxies, we find that  
while in the first three radio flux bins the fraction of early type 
galaxies is approximately constant at $\sim$40\%, 
in the fainter radio flux bin this fraction decreases to 
$\sim30$\%.  Therefore, although this result is still highly qualitative and based 
on very simple assumptions, we confirm that in the fainter radio flux limit bin  
our radio sample  is likely to be  dominated by late-type star forming galaxies as already suggested by 
the analysis of the optical magnitude distribution in different radio flux intervals 
(see Section  4.1). 

%%%%%%%%%%%%%%%%%%%%%%%%%%%%%%%
%  Color - redshift diagram
%%%%%%%%%%%%%%%%%%%%%%%%%%%%%%%

%%%%%%%%%%%%%%%%%%%%%%%%%%%%%%%
% End figure
%%%%%%%%%%%%%%%%%%%%%%%%%%%%%%%

%%%%%%%%%%%%%%%%%%%%%%%%%%%%%%%
%  Radio to optical ratio for early and late 
%%%%%%%%%%%%%%%%%%%%%%%%%%%%%%%

%%%%%%%%%%%%%%%%%%%%%%%%%%%%%%%
% End figure
%%%%%%%%%%%%%%%%%%%%%%%%%%%%%%%

\subsection{Near Infrared magnitude properties} 

\begin{figure*}
\centering

\psfig{figure=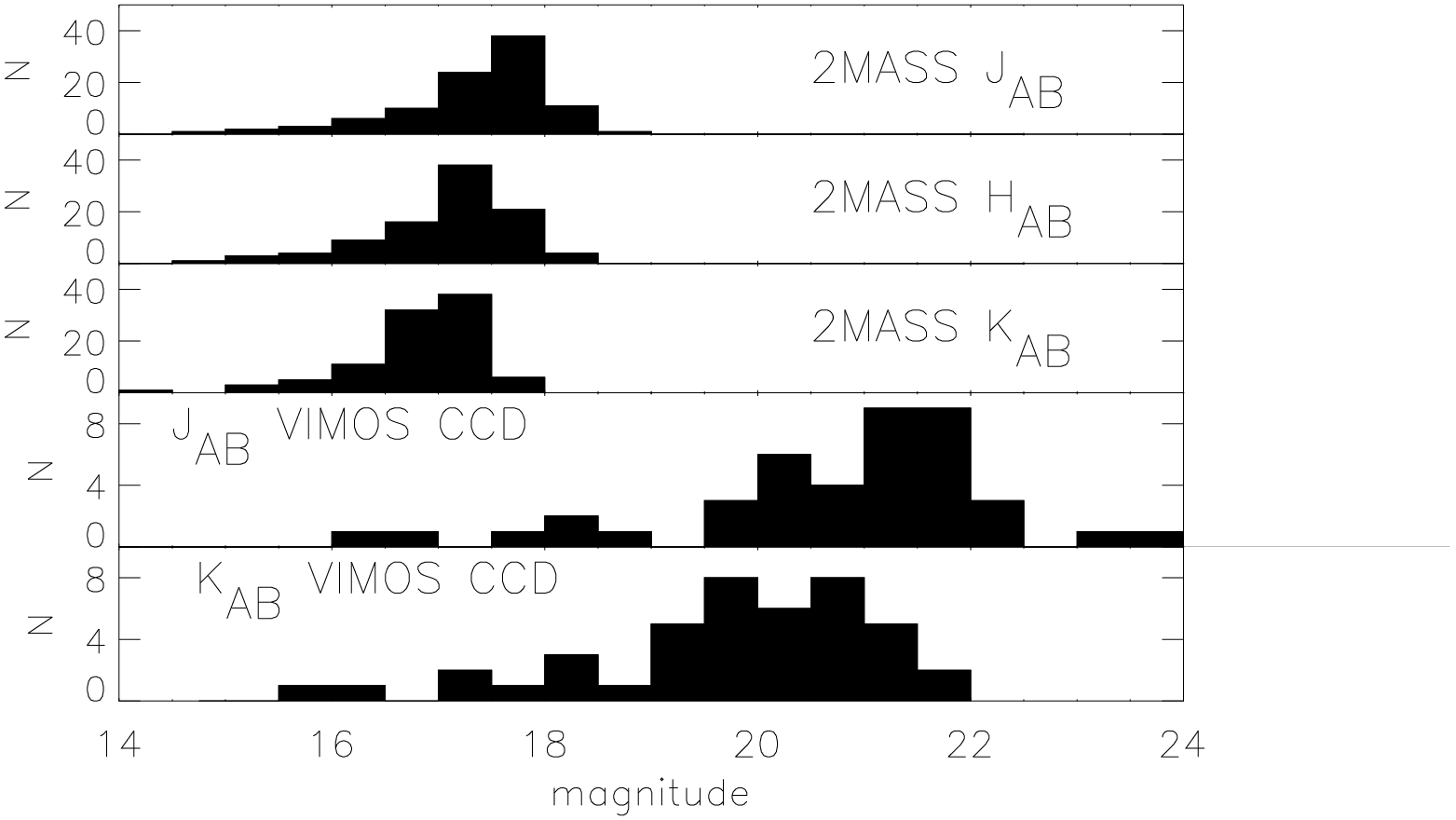,width=15cm}

\caption[]{The magnitude distributions in the three 2MASS near infrared bands 
for the 105 radio sources with a reliable conunterpart in the 2MASS catalogue 
and the magnitude distributions in the $J_{AB}$ and $K_{AB}$ bands for the near 
infrared counterparts of the 43 radio sources identified in the VIMOS near infrared 
survey (bottom panels). 
The 2MASS $J,H,$ and $K$ magnitudes have been converted to the 
AB magnitude system using the relations reported in Section 3.4. }

\label{NIR_histo}

    \end{figure*}

\begin{figure*}

\centering

\psfig{figure=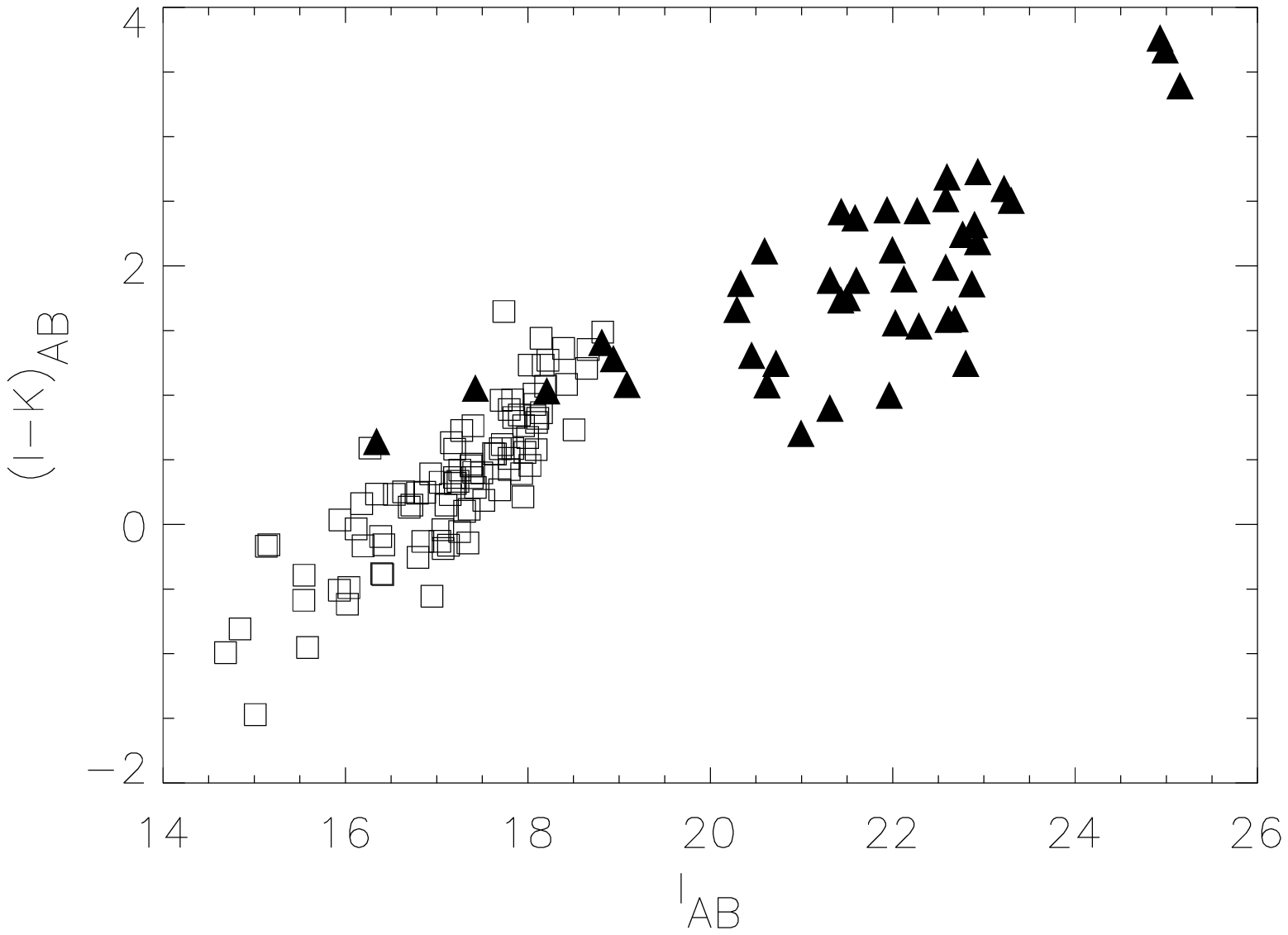,width=15cm}

\caption[]{The  $(I-K)_{AB}$ color as function 
of the $I_{AB}$ magnitude. Open squares are sources from 
the 2MASS data while filled triangles are sources from the 
VIMOS $K$ band survey.} 

\label{IK_I}

    \end{figure*}

In Figure~\ref{NIR_histo}  we report the magnitude 
distributions in the three 2MASS bands for the 105 sources with an 
identification in the 2MASS catalogue and the  J$_{AB}$ and 
K$_{AB}$ distributions 
for the 43 radio sources identified with the VIMOS near infrared  survey. 
Moreover  Figure~\ref{IK_I} shows the $(I-K)_{AB}$ color as function 
of the $I_{AB}$ magnitude. In considering these figures, it must be 
remembered how incomplete these data are : only 65 of the radio sources 
are within the area covered by the VVDS J and K bands survey, while the 
radio  flux distribution of the radio sources with a 2MASS counterpart 
is  strongly biased towards high radio flux due to the bright limit of 
the 2MASS survey. Figure~\ref{IK_I} clearly shows that there is a 
monotonic trend for the optically fainter radio sources to have 
redder optical-infrared colors, and their color reaches $(I-K)_{AB}\geq$2 at 
$I_{AB}\sim$23 mag. Similar results have been obtained for different 
radio surveys (Richards et al. 1999, Waddington et al. 2000, Ciliegi et al. 
2003).   From  Figure~\ref{IK_I} it is also interesting to note that 
only 7\% (3/43) of the radio sources are very red sources with  $(I-K)_{AB}>$3.0.
An even smaller percentage of EROS has been recently obtained by Sullivan et al. 2004 during the
optical and near infrared identification of the radio sources in the 
Phoenix Deep Survey.  None of the 91 radio sources detected in the
$K$  band (over a total of 111 radio sources)  
 has an  $(I-K)_{AB}$ color greater  than 3.0 and only one source has 
 an  $(I-K)_{AB}>$2.5.  The fact that EROs sources are only a few percent 
of the total number of counterparts in 
radio surveys  with a flux limit  around $\sim$100$\mu$Jy 
is not surprising. Deep radio observations of optically/near-infrared 
selected EROS have shown that the EROs population is not a class of strong radio 
sources.  For example, Smail et al. 2002, using very deep 1.4GHz radio 
data (3.5$\mu$Jy at 1$\sigma$) found only 3 EROs (over a total of 68) 
with a radio flux greater than 85$\mu$Jy, while 
Cimatti et al. 2003 studying a sample of 47 EROs,  detected only one object at 1.4 GHz 
with a radio flux of $\sim$106 $\mu$Jy. However, the correlation between color and 
magnitude shown in Figure~\ref{IK_I} suggests that the fraction of EROs is likely to be higher 
among the unidentified radio sources. This can be tested with deeper optical and 
infrared data.

%%%%%%%%%%%%%%%%%%%%%%%%%%%%%%%
% Near infrared Magnitude distribution figure
%%%%%%%%%%%%%%%%%%%%%%%%%%%%%%%
%%%%%%%%%%%%%%%%%%%%%%%%%%%%%%%
% End figure
%%%%%%%%%%%%%%%%%%%%%%%%%%%%%%%

%%%%%%%%%%%%%%%%%%%%%%%%%%%%%%%
% (I-K) vs I  figure
%%%%%%%%%%%%%%%%%%%%%%%%%%%%%%%
%%%%%%%%%%%%%%%%%%%%%%%%%%%%%%%
% End figure
%%%%%%%%%%%%%%%%%%%%%%%%%%%%%%%

\section{Summary and Conclusion} 

In this paper we have presented the optical and near-infrared identifications 
of the 1054 radio sources detected in the 20cm deep radio survey obtained 
with the VLA in the VIMOS VLT Deep Survey VVDS-02h deep field (VVDS). Almost the whole square 
degree of the VVDS-VLA field has been observed in the $B,V,R$ and $I$ band 
down to a limiting magnitude of  $B_{AB}\sim$26.5,
$V_{AB}\sim$26.2,  $R_{AB}\sim$25.9 and   $I_{AB}\sim$25.0 
(McCracken et al. 2003).  Moreover, 
 $\sim$0.71 deg$^2$ have been observed in the $U$ band down to 
$U_{AB}\sim$25.4 (Radovich et al. 2004) 
and $\sim$165  arcmin$^2$ have been observed in 
the $J$ and $K$ bands down to $J_{AB}\sim$24.2 and $K_{AB}\sim$23.9 (Iovino et al. 2005).   
Using the Likelihood Ratio technique we optically identified 
718 radio sources ($\sim$74\% of the whole sample) . Sixty-five radio sources 
lie within the $K$ band area and  we found a reliable counterpart for 43 of them. 
Among the 255 unidentified radio sources, 17 are empty fields ($i.e.$ 
they have no optical source within 5 arcsec from their position), while 
the other 238  sources have at least one optical source within 5 arcsec but 
all with $LR < LR_{\rm th}$.   

The  color properties of the optical counterparts of the radio 
sources have been 
analysed  using the $(B-V)_{AB}$ and $(V-I)_{AB}$ colors.  The optical 
counterparts of the 
radio sources classified as extended have been analysed in 
three different magnitude 
slices. While in the brightest magnitude range  (18$<I_{AB}<$20) the optical 
color properties of the radio sources are not significantly different 
from those of the whole 
optical sample, at fainter magnitude the median $(V-I)_{AB}$ color of the 
radio sources 
is redder than the median color of the whole optical sample, suggesting 
a higher redshift for the radio sources. This  is also supported 
by the photometric redshift analysis which shows that, in each magnitude
bin, the radio sample has a higher median photometric redshift than 
the whole optical sample. This suggests that radio detection is preferentially 
selecting galaxies with higher intrinsic optical luminosity. 

Using the $(B-I)_{AB}$ color and the photometric redshift for all the  radio sources 
with a reliable optical counterpart, we have tentatively divided the radio sample 
in two sub-samples: 
the sources above the Sa galaxies track have been considered as 
early type galaxies, while the sources below the same track have been considered as 
late type galaxies.  The analysis of the radio-to-optical ratio $R$ of the two 
sub-samples confirms  with high statistical significance 
the results already obtained by other authors :   late-type, star forming 
galaxies are the dominant population at low $R$, while in the region of 
high $R$ the majority of these objects are early-type galaxies. 

From the analysis of the optical properties of the radio sources in different 
radio flux bins,  we found that while about 35\% of the radio sources are 
optically unidentified in the first  radio flux bin, the percentage 
of unidentified sources decreases to about 25\% in the faintest two bins 
(S$<$ 0.5 mJy). The median $I_{AB}$ magnitude for the total sample of radio sources, 
i.e. including also the unidentified ones, is brighter in the faintest 
two radio bins than in the bin with higher radio flux.  This 
result shows that the faintest radio sources are not in general the faintest 
sources at optical wavelengths and would 
suggest  that 
most of the faintest radio sources are likely to be associated to relatively
lower radio luminosity objects at relatively modest redshift, rather 
than radio-powerful, AGN type objects at high redshift. Using 
the above classification in early-type  and late-type galaxies  we found 
that the majority of the radio sources below $\sim$0.15 mJy are indeed late-type star forming 
galaxies in the photometric redshift range 0.1$\leq$z$_{phot}\leq$1.5.  
These results are in agreement with the results obtained by 
several authors : the majority of the 
optical identification of the $\mu$Jy radio sources are with luminous ($L>L_*$) galaxies 
at modest redshifts (0$\leq$z$\leq$1), many of which with evidence for recent star formation
(Windhorst et al. 1995, Richards et al. 1998, Richards et al. 1999, 
Roche et al. 2002). 

Finally, as already noted by other authors for different radio surveys, 
we found a monotonic trend for the optically fainter radio sources to be 
associated with redder galaxies.  In the area covered by $K$ data 3 out of 43 
radio sources with a likely $K$ band counterpart have very red $(I-K)_{AB}$ colors. 

The analysis of the sub-sample of spectroscopically identified 
radio-sources is in progress and will be presented elsewhere.

\begin{acknowledgements}
 
This research has been developed within the framework of the VVDS consortium.\\
This work has been partially supported by the Italian Ministry (MIUR) grants
COFIN2003 (num.2003020150).\\

This publication makes use of data products from the Two Micron All Sky Survey, which is a 
joint project of the 
University of Massachusetts and the Infrared Processing and Analysis Center/California
 Institute of Technology, 
funded by the National Aeronautics and Space Administration and the National Science Foundation. 

\end{acknowledgements}

\end{document}